# EUROPEAN LONGITUDE PRIZES. II. ASTRONOMY, RELIGION AND ENGINEERING SOLUTIONS IN THE DUTCH REPUBLIC

**Richard de Grijs**
*Department of Physics and Astronomy, Macquarie University,*
*Balaclava Road, Sydney, NSW 2109, Australia*
Email: richard.de-grijs@mq.edu.au

**Abstract:** The late-sixteenth century witnessed a major expansion of Dutch shipping activity from northern European waters to the Indian Ocean and beyond. At a time when the Renaissance had just arrived on the North Sea's shores, scientist-scholars, navigators and merchants alike realised the urgent need for and potential profitability of developing a practical means of longitude determination at sea. Under pressure of early adopters, including Petrus Plancius and Simon Stevin, on 1 April 1600 the national government of the Dutch Republic announced a generous longitude prize, which would see gradual increases in value over the next two centuries. In addition to leading thinkers like Galileo and Christiaan Huygens, the Low Countries spawned major talent in pursuit of a longitude solution. Their solutions reached well beyond applications of the ephemerides of Jupiter's moons or the development of a stable marine timepiece. Studies of the Earth's magnetic field, lunar distances, astronomical observations combined with simple trigonometry and the design of a 'golden compass' all pushed the nation's maritime capabilities to a higher level. Dutch efforts to 'find East and West' were unparalleled and at least as insightful as those pursued elsewhere on the continent.

## 1 EXPANDING TRADE ROUTES

Although international trade networks and shipping routes saw tremendous growth in late-medieval times, in Europe that expansion was initially almost entirely driven by the Spanish (Castilian)–Portuguese rivalry and the associated quest for maritime dominance (for a review, see de Grijs, 2020*b*). By the 1580s, however, sailors hailing from Flemish ports and the Hanseatic cities extended their active range beyond North and Baltic Sea coasts to destinations as far away as the East Indies. They thus required more sophisticated navigational aids to safely travel across the open oceans (Schilder and van Egmond, 2007). Over the course of the next few centuries, this pressing need led directly to the establishment of well-funded longitude prizes (e.g., Howse, 1998) by the governments of Castile ('Spain'), the Dutch Republic, the Venetian Republic and Great Britain, as well as by the French *Académie Royale des Sciences* through its Meslay Prize (e.g., Boistel, 2015).

King Philip II, the Spanish monarch, established a generous longitude prize in 1567, which his son, Philip III, confirmed and increased in value upon his accession to the Spanish throne in 1598 (de Grijs, 2020*b*). The *Staten Generaal* (States General) of the United Provinces of the Dutch Republic, and to a lesser extent also lower-level governing bodies (van Berkel, 1998), had been under pressure since the early 1590s by scientists—including Petrus Plancius, Simon Stevin and Matthijs Lakeman (Davids, 1986: 69; Wepster, 2000)—to issue their own reward to anyone who could provide an adequate and practical solution to what the Dutch referred to as the problem of "finding East and West".

On 1 April 1600, the States General finally announced the availability of a reward of 5,000 carolus guilders (*florins*), as well as a life annuity of 1,000 pounds, for successful applicants (Davids, 2009):

> **346. Resolution 1 April** – At the request of Jacob van Straten,[1] … [it] has been resolved that, while there are several others who have indicated that they have found the same [a method to determine longitude at sea], that all contenders will be required to provide written explanations of their inventions within six weeks, or at most two months, upon which all written submissions will be opened and examined, that it is promised and agreed that to those who have offered a genuine invention against which there are no objections, an annual sum of one thousand pounds of twenty *stuivers*[2] each and five thousand guilders in cash … will be paid, inasmuch … that if multiple [submissions] have resulted in perfect inventions, that [the successful contenders] will share their written explanations with the others and compete. (Davidse, 2000–2020)

This resolution was well overdue. At least five contenders had entered the fray even before the first Dutch longitude prize had been established (Davids, 1986: 69). There was a clear need for facilitation and encouragement of navigational progress by the highest authorities. Although there was no direct link to the Spanish longitude competition, a number of submissions originated from applicants who had been inspired by and often had first submitted their proposals to the Spanish *Casa de la Contratación* (House of Trade) in Seville (Spain). As we will see below, Galileo Galilei[3] was a 'repeat submitter', although his proposal to the Dutch authorities was not necessarily identical to that submitted to the Spanish Crown.

On 7 July 1611, on the recommendation of Hendrick Willemsz Nobel, Albert Joachimi and Willem[4] van Velsen (councillors and expert assessors), the States General increased its prize amount. The reward was consolidated into a single, maximum lump sum of 15,000 guilders (Dodt van Flensburg, 1846: 244; Davidse, 2000–2020). In 1660 it was increased once again, to 25,000 guilders. A year after the States General had announced their initial reward, the *Staten van Hollant ende Westfrieslant* (the provincial States of Holland and Westfrisia) announced their own version of a longitude prize, offering 150 guilders to anyone for an initial written explanation of their method. This prize would be offered provided that the proposers were prepared to have their methods tested at sea. In addition, a lump sum of 3,000 guilders and an annuity of 1,000 guilders would be awarded if six to eight navigation experts would attest to the method's practicality and reliability (de Grijs, 2020*b*). By 1738, the States of Holland were prepared to pay as much as 50,000 guilders to successful applicants.

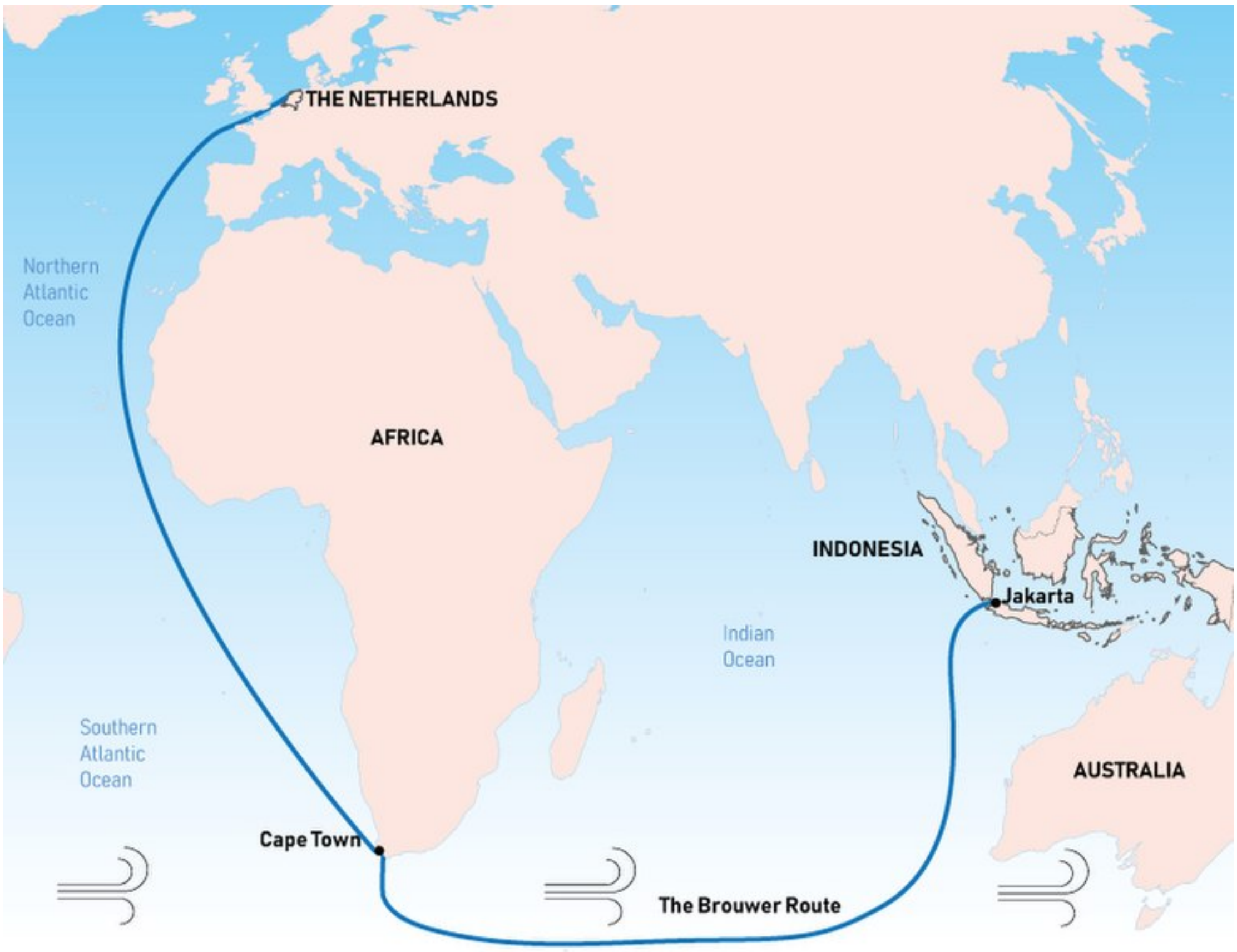

**Figure 1**: Map of the 'Brouwer' route. (Courtesy: Redgeographics, via Wikimedia Commons; Creative Commons Attribution-Share Alike 4.0 International license)

The gradual increases in the prize amounts on offer may have been related to changes in the shipping routes to the Dutch East Indies. Initially, ships commandeered by the *Vereenigde Oostindische Compagnie* (VOC; the Dutch East India Company, established in 1602) adopted direct routes across the Indian Ocean from the Cape of Good Hope to Batavia (present-day Jakarta, Indonesia). However, from 1616 a different course was mandated, which involved a rapid transition across the Indian Ocean at a latitude of about 40° South, followed by a directional change to the north at the approximate longitude of Batavia (Wepster, 2000): see Figure 1. This southern, 'Brouwer' route took advantage of the prevailing westerly trade winds ('the roaring forties'), thus allowing completion of the voyage much more rapidly than before. Meanwhile, the more temperate climate contributed to a better overall health profile of the crews and a lower incidence of perishable-food decay.

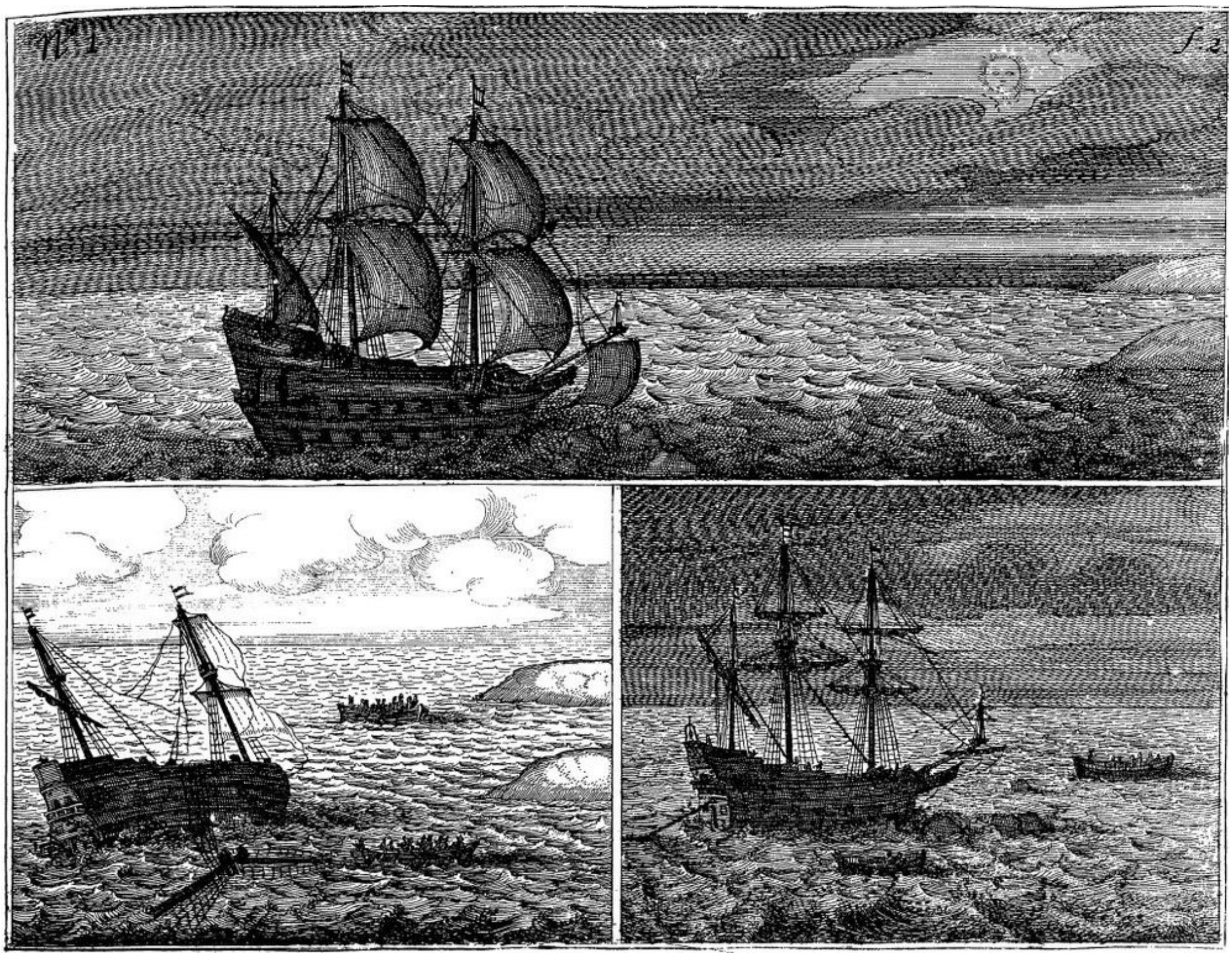

**Figure 2**: Plate 1 from *Ongeluckige voyagie, van't schip Batavia* (*Unlucky voyage of the ship Batavia*; 1647). (Wikimedia Commons; public domain)

Determination of one's longitude while on the rapid southerly leg was of paramount importance. A premature turn northwards would direct the ships to the Sumatra coast, from whence sailing conditions across the Sunda Strait to Batavia were treacherous. A late course change, however, could be even more devastating, as evidenced for instance by the Dutch East Indiaman *Batavia*'s shipwreck (see Figure 2) on the Houtman Abrolhos off the Western Australian coast in 1629.

Eventually, by 1775, almost forty proposals had been submitted for one of the Dutch longitude rewards. Most applicants were based in the Dutch Republic, although a number of high-profile hopefuls—including Galileo, in 1634—hailed from abroad, particularly from the Holy Roman Empire and from France (Davids, 2009). Many were philosophers, scientists or those employed in affiliated professions (e.g., Andreas van Berlicom in the mid-seventeenth century; van Berlicum, 1656). Perhaps surprisingly, only few were practicing sailors (Wepster, 2000). Yet, despite this flurry of proposals, neither the States General nor the States of Holland ever awarded the full prize money. Yet, they regularly offered promising contenders significant expense allowances. Among those provided with an expense budget, the most successful contender was Jan Hendricksz Jarichs van der Ley (see Figure 3[5]). In 1625, he was awarded an annuity of 1,200 guiders for himself and his heirs, which would pay out 19,000 guilders by the time the agreement was cancelled in 1655 (Davids, 2005).

In Spain, all proposals vying for a share of its longitude prize were assessed by a single agency, the *Casa de la Contratación*, acting through the *Conseja de Indias* (Council of the Indies). In the Dutch Republic, however, the initial assessments of and responses to the various proposals received by the governing bodies were handled in a more *ad hoc* manner, involving a larger number of stakeholders covering a wider range of expertise. Most commonly, the States General would convene an *ad hoc* committee of *theoristen*, experts in the theory of navigation, to solicit their expert opinion as to the theoretical correctness of the proposed solution. If a proposal passed that initial assessment, '*practisijns ende stierluijden*', experienced sailors (literally, practicing navigators), would be enlisted to provide their expert advice as to whether the solution was "completely secure and certain" (Davids, 1986: 73) and also practically viable. This penultimate phase might be repeated a number of times, sometimes also including further discussions and debate regarding the method's merits. The final assessment phase then involved actual sea trials.

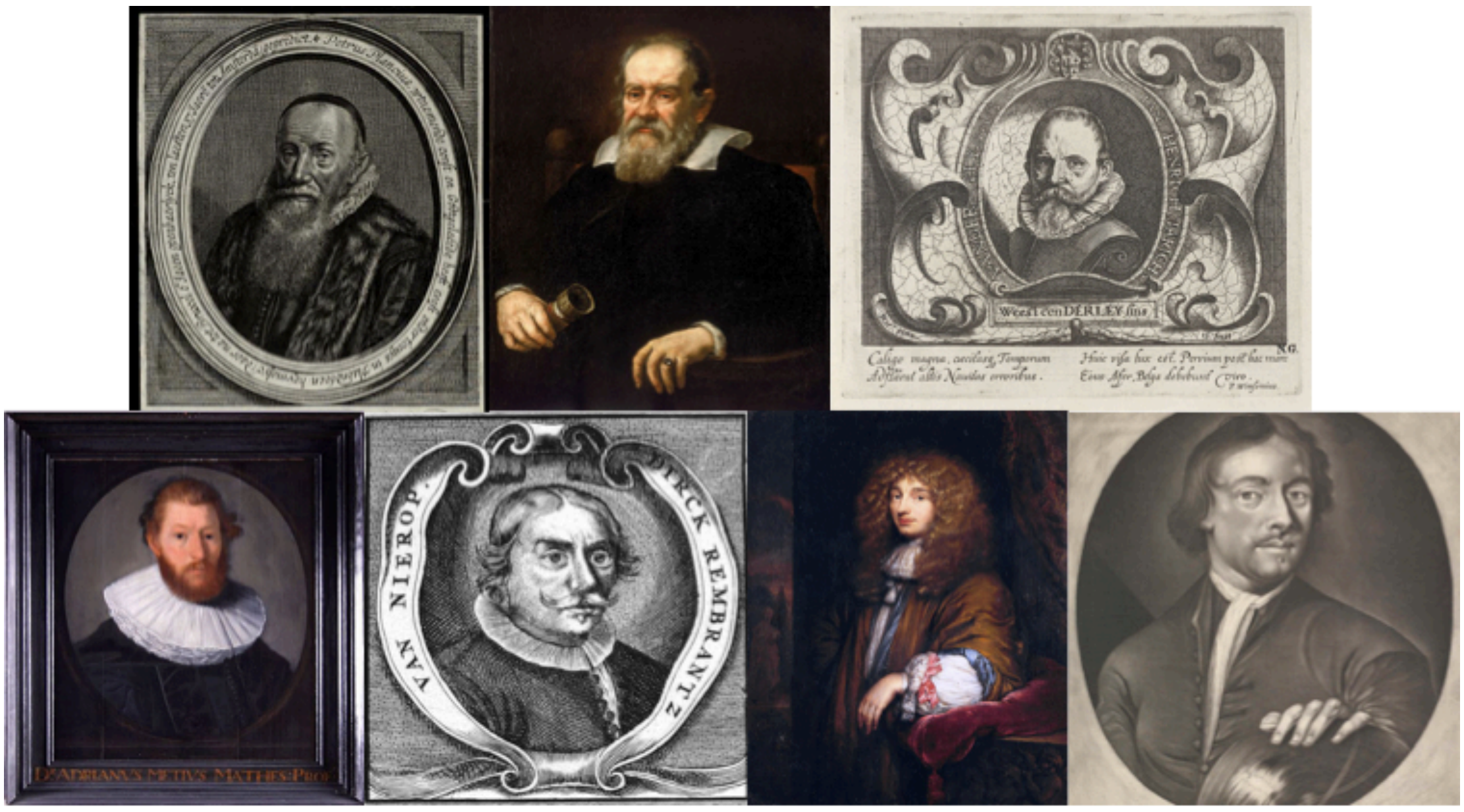

**Figure 3**: Portraits of the main characters driving the innovations described in this article, both scientist-scholars and opportunistic 'projectors', ordered from left to right and from top to bottom by date or birth. Individuals depicted include *(i)* Plancius (1552–1622); *(ii)* Galileo (1564–1642); *(iii)* Jarichs van der Ley (c. 1566–1639); *(iv)* Metius (1571–1635); *(v)* van Nierop (1610–1682); *(vi)* Christiaan Huygens (1629–1695); and *(vii)* Willemsz (Graaf) (1652–1704).
*Figure credits*: Wikimedia Commons, except for *(vii)*: Rijksmuseum Amsterdam. All images are in the public domain, except for *(iii)*: Creative Commons CC0 1.0 Universal Public Domain Dedication.

The *ad hoc* committees of *theoristen* included established scholars, teachers and surveyors (see Figure 4[6])—including Joseph Justus Scaliger, Rudolf and Willebrord Snellius, Stevin, Sybrand Hansz Cardinael (Sybrand Hanssen), Franciscus (Frans) van Schooten, Christiaan Huygens, Martinus Hortensius (van den Hove), Burchard de Volder, Petrus (Pieter) van Musschenbroek, Willem Jansz Blaeu and Isaac Beeckman, among others. Initially, the States General would invite expert commentary by letter, although oral reports to the full assembly were also acceptable. This process became more formalised by the middle of the seventeenth century. The government would pass a resolution to request the expertise of earmarked technical experts, who would then be sent a formal missive, accompanied by any relevant material for assessment (Thomassen, 2009). Reports had to be delivered in written form, a practice implemented from the 1630s.

Many of these *theoristen* committees drew heavily on the expertise of teachers and scholars at universities or prominent ('*Illustre*') colleges (Davids, 1990). For instance, two Leiden University employees, including Rudolf Snellius, professor of mathematics since 1581, and Scaliger, professor of history since 1591, were appointed to one such committee in 1598, together with Stevin, Lucas Jansz Waghenaer and Ludolf van Ceulen (Davidse, 2000–2020). On 3 November 1617 the States General requested that Stevin, Gerard Meerman, Jacob Magnus and Bocko van Burmania assess Jarichs van der Ley's proposal. At various times between 1618 and 1620, Adriaen Metius, Nicolaus Mulerius and Willebrord Snellius, as well as a number of surveyors, experienced sailors and instructors of navigation were also invited to offer their assessments on Jarichs van der Ley's proposal (Davids, 1986: 80–82, 284–287). Meanwhile, in 1635 van den Hove, Blaeu and Beeckman were asked "to receive a written description of and examine" Galileo's proposal (Davidse, 2000–2020; Thomassen, 2009). I will discuss these proposals in detail below.

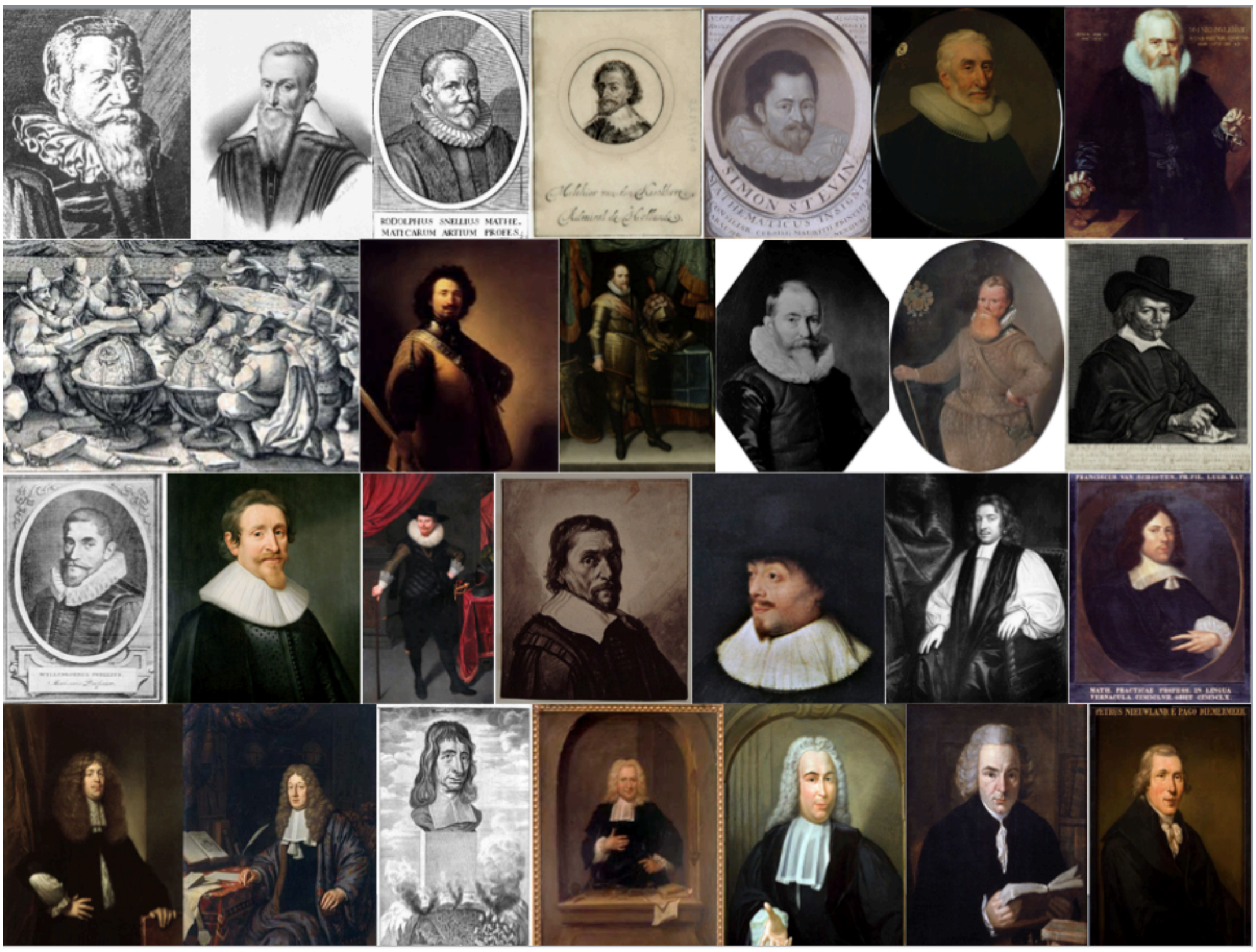

**Figure 4**: As Figure 3 but for the officials, examiners and support personnel playing important roles in the validation of all longitude solutions discussed in this article. *(i)* Ludolf van Ceulen (1540-1610); *(ii)* Scaliger (1540–1609); *(iii)* Rudolf Snellius (1546–1613); *(iv)* van den Kerckhove (d. 1619); *(v)* Stevin (1548–1620); *(vi)* Jan (Johan) de Groot (1554–1640); *(vii)* Mulerius (1564–1630); *(viii)* le Canu (1563–1630), teaching a nautical class; *(ix)* Carolus (c. 1566–c. 1636); *(x)* Prince Maurits (1567–1625); *(xi)* Blaeu (1571–1638); *(xii)* de Houtman (1571–1627); *(xiii)* Dou (1572–1635); *(xiv)* Willebrord Snellius (1580–1626); *(xv)* Hugo de Groot (1583–1645); *(xvi)* Reael (1583–1637); *(xvii)* Colvius (1594–1671); *(xviii)* Constantijn Huygens (1596–1687); *(xix)* Hartlib (c. 1600–1662); *(xx)* van Schooten (1615–1660); *(xxi)* van Beuningen (1622–1693); *(xxii)* Hudde (1628–1704); *(xxiii)* Bekker (1634–1698); *(xxiv)* van Musschenbroek (1692–1761); *(xxv)* Lulofs (1711–1768); *(xvi)* van Swinden (1746–1823); and *(xvii)* Nieuwland (1764–1794). *Figure credits*: Wikimedia Commons, except for *(iv)*: Regional Archives Dordrecht; *(viii)*: Oudhoorn Archives; and *(ix)*: *Wanna Know History* blog, November 2015. All images are in the public domain, except for *(vi)*: Creative Commons CC0 1.0 Universal Public Domain Dedication; *(vii)*: Creative Commons Attribution-Share Alike 3.0 Unported, 2.5 Generic, 2.0 Generic and 1.0 Generic licenses; and *(xvii)*: Creative Commons Attribution-Share Alike 2.5 Generic license.

In the second half of the seventeenth century, this external expertise was still highly sought after. On several occasions in the 1680s and 1690s, de Volder—professor of physics, astronomy and mathematics at Leiden University—developed into a high-profile expert providing support to the national government, the provincial government of Holland, as well as the Amsterdam chamber of the VOC in their assessment of newly proposed longitude solutions (Davids, 1986: 132, 135–137). In fact, the VOC became a centre of expertise of sorts, given that it employed dedicated pilot examiners, who were experts in navigation themselves. As such, from the middle of the seventeenth century until about 1730, the States General frequently referred contenders for its longitude prize to the VOC (Davids, 1986: 73, 81–83, 132, 180; Vanpaemel, 1989).

The joint pursuit of trade and practical science came naturally to the sailors engaged in the East India voyages and their paymasters. Scientific endeavours were pursued systematically ever since the first Dutch voyage to Asia in 1595 (see Figure 5), the so-called *Eerste Schipvaart* (van Berkel, 1998). On that voyage, Plancius took charge to obtain sufficient numbers of observations of variations of the 'magnetic declination' (deviations of the compass needle from true North) as a potential but ultimately unsuccessful means to determine one's longitude at sea.

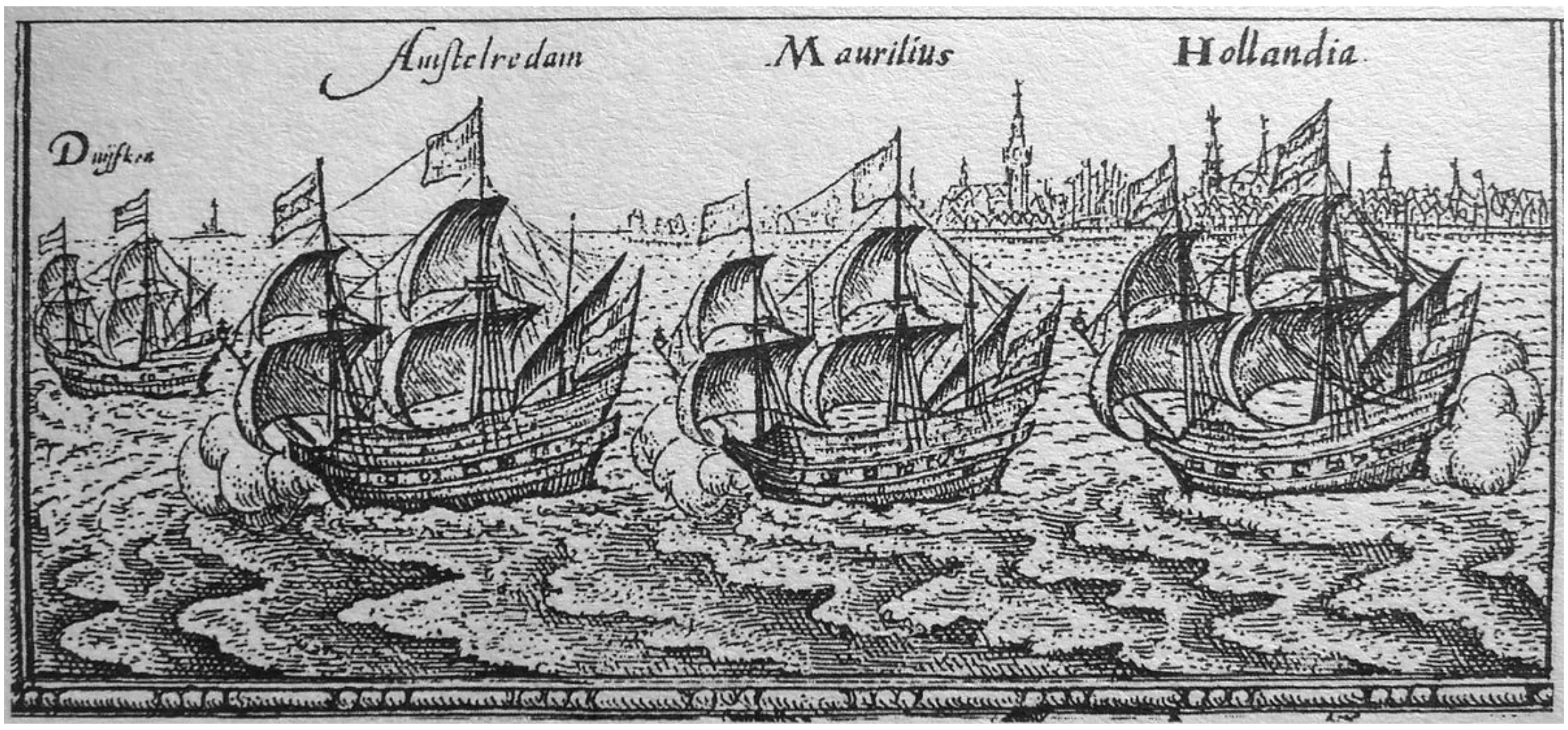

**Figure 5**: The Dutch East India fleet of the *Eerste Schipvaart* (1595–1596), commanded by Cornelis de Houtman. (Source: Willard Hanna, Bali Chronicle, via Wikimedia Commons; public domain)

From the 1740s, the States General increasingly relied on the services of technical experts affiliated with universities and *Illustre* colleges, which was reflected in their job descriptions. Between 1743 and 1796, lecturers in mathematics, astronomy and nautical science at Amsterdam's *Athenaeum Illustre* (see Figure 6)—the predecessor of the University of Amsterdam—were jointly appointed as pilot examiners at the VOC's Amsterdam chamber. Similarly, the VOC's pilot examiner at its Zeeland chamber in Middelburg was simultaneously appointed as professor of philosophy, mathematics, physics and astronomy at the town's *Illustre* college.

In 1787, the Admiralties of Amsterdam and Rotterdam formalised these arrangements by creating the Committee on Matters Relating to the Determination of Longitude at Sea and the Improvement of Charts[7] (Davids, 2015), which became an agency of the Dutch Navy from 1795 until its dissolution in 1850. This Longitude Committee always included one or two professors from either the *Athenaeum Illustre* in Amsterdam, Leiden University or Utrecht University. Its main purpose was to disseminate knowledge regarding maritime cartography, longitude determination at sea and other aspects of nautical science (Davids, 1986: 188, 342, 399–400; Davids, 2016). One of its most important tasks was the translation and adaptation of the British *Nautical Almanac* for a reference meridian through Tenerife's Teide volcano instead of Greenwich.

Only a small fraction of the initial submissions eventually reached the sea trial phase, and if they did, prolonged debates about their viability often ensued upon the ships' return. Among the most promising proposals whose development was taken forward to practical sea trials (Davids, 2008: 446) was Jarichs van der Ley's improved technique of 'dead reckoning', which was taken on a "voyage of the experiment" in 1618. Later that century, a number of Huygens' marine timepieces were tested on sea trials undertaken as part of regular VOC voyages to the Cape of Good Hope in the 1680s and 1690s (e.g., de Grijs, 2017). In the 1730s, the VOC also undertook a number of sea trials of instruments for the improved measurement of speed and leeway invented by Leendert Vermase and Jasper van der Mast.

From approximately the 1580s, Dutch knowledge of maritime navigation had improved rapidly, allowing the nation's scholars to become leaders in marine cartography, particularly of the North and Baltic Seas. The most famous sixteenth-century cartographer was Waghenaer, on account of his publication of the first illustrated book of sailing directions, *Spieghel der Zeevaerdt* (*Mariner's Mirror*; 1584–1585: see Figure 7). However, in the modern public mind, the history of longitude determination in the Netherlands is firmly associated with Huygens' pendulum clocks and, to a lesser extent, Galileo's unsuccessful attempts at securing one of the Dutch longitude prizes. Yet, well before the heyday of these illustrious scholars, from the sixteenth century, the nation's scientists and navigators were actively engaged in practical approaches to solving this most vexing problem. It is my aim to highlight the advances made in this field over the better part of two centuries, by providing a comprehensive overview of the people, methods and developments involved and pursued.

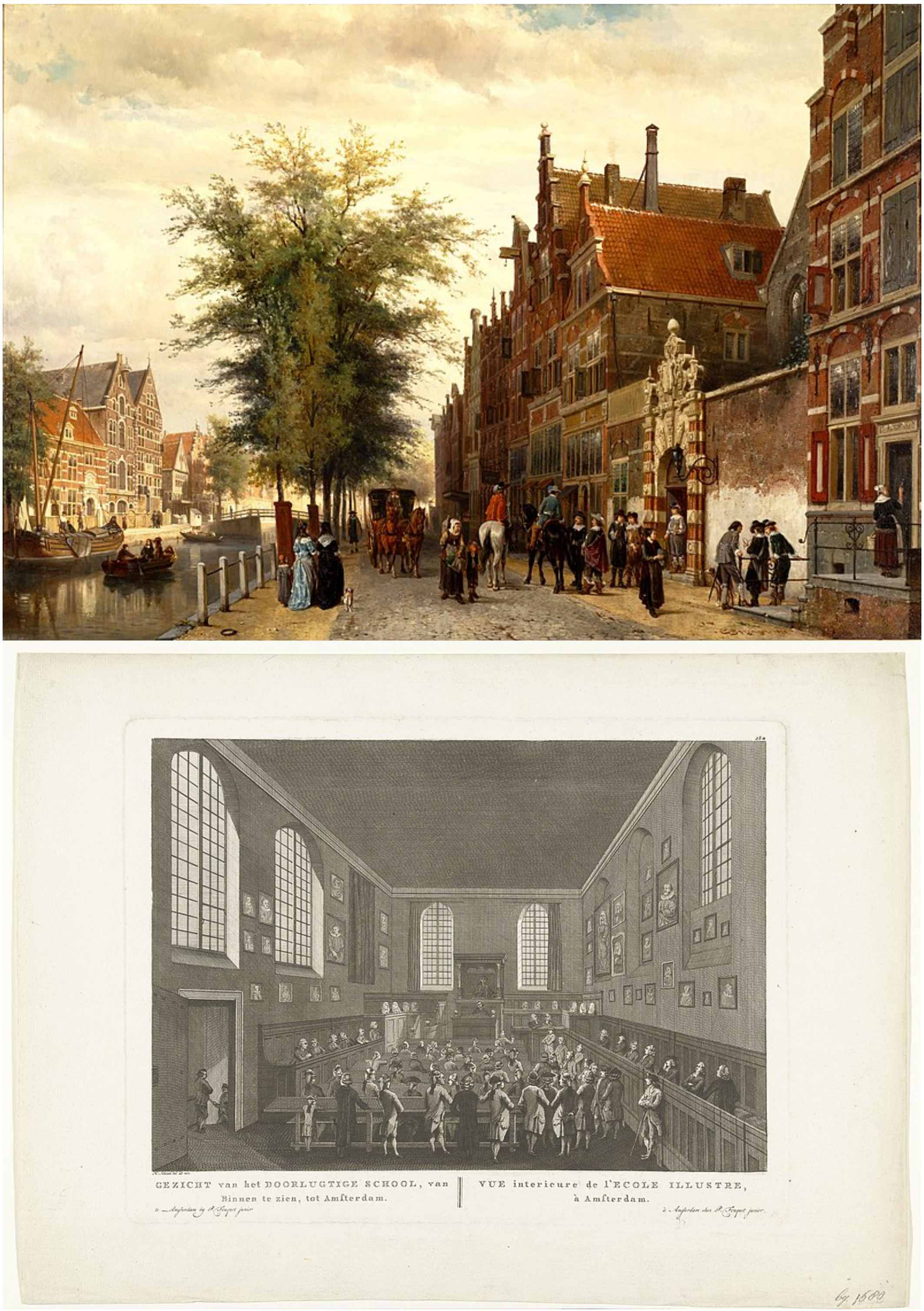

**Figure 6**: (*top*) The Amsterdam *Athenaeum Illustre* in 1650 (*Cornelis Springer, 1878).* (Courtesy: Teylers Museum, Haarlem, via Wikimedia Commons; public domain); (*bottom*) Interior (Hermanus Petrus Schouten, c. 1770–1783). (Courtesy: Rijksmuseum Amsterdam, via Wikimedia Commons; Creative Commons CC0 1.0 Universal Public Domain Dedication)

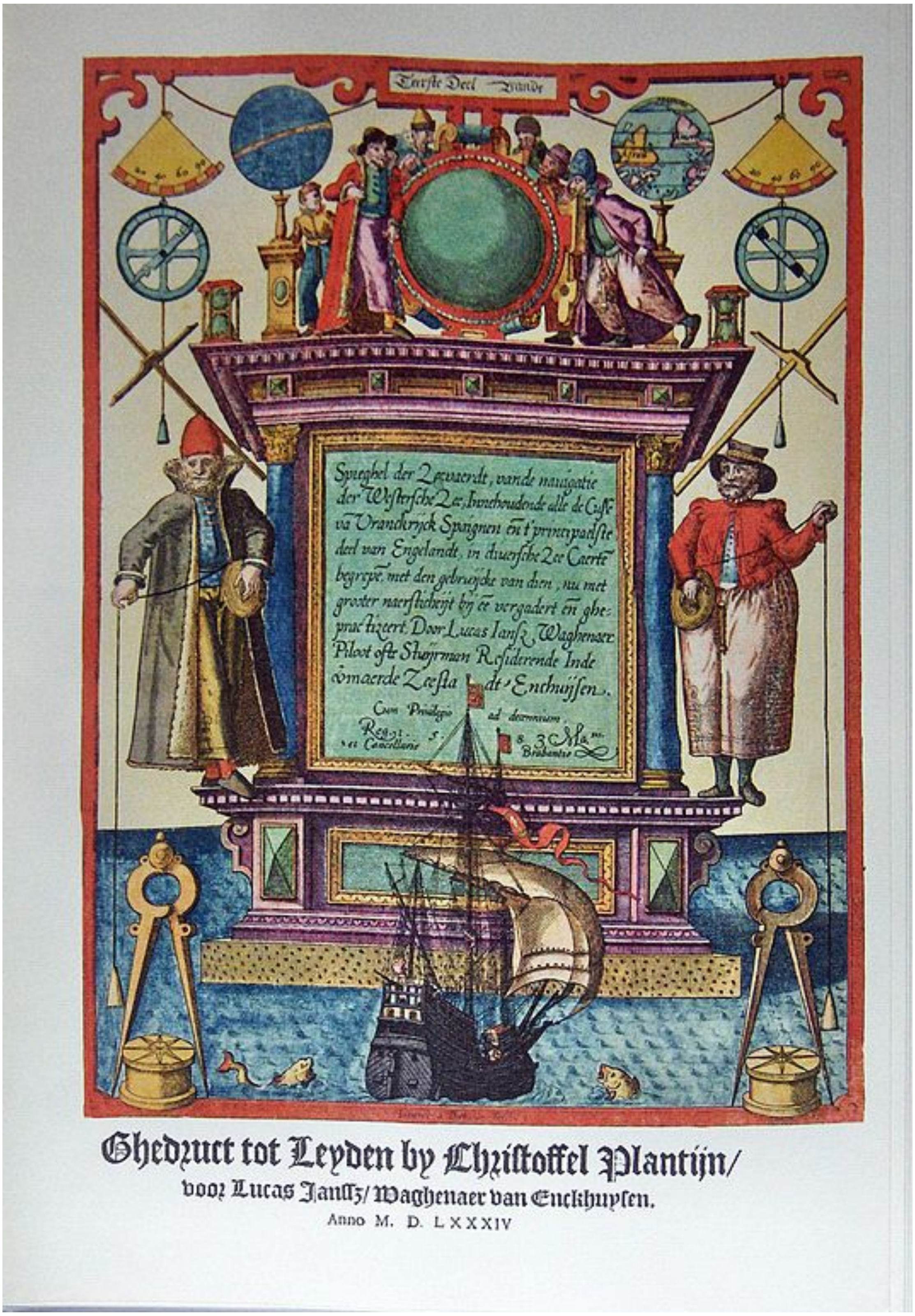

**Figure 7**: *Spieghel der Zeevaerdt* (title page), a collection of coastal maps from 1584–1585 by Lucas Jansz Waghenaer. (Wikimedia Commons; Creative Commons Attribution 3.0 Unported license)

## 2 EARLY CONTENDERS

The collection of resolutions passed by the States General around the turn of the sixteenth century represents a treasure trove of information about early efforts to develop novel methods for longitude determination at sea. These resolutions, which have been made available online[8] for historical research, encompass a complete record of any such methods offered to the Dutch government for possible patenting between 1576 and 1625.

### 2.1 Patent applications leading up to the Dutch prize announcement

Not surprisingly, the earliest of the resolutions referring to a new longitude invention relates to a prominent contemporary scholar: Petrus Plancius. This Dutch–Flemish astronomer and cartographer was a Reformed Church clergyman from the southern Netherlands. He followed in the footsteps of his illustrious Spanish predecessors by whom he was influenced—including João de Lisboa, Francisco Faleiro and Alonso de Santa Cruz—by proposing a possible solution based on observations of the magnetic compass needle and its variation with position (Davids, 1986: 70–71, 75; Jonkers, 2000).

Plancius was convinced that a global pattern showing the relationship between magnetic declination and the direction of true North could be derived (van Berkel, 1998). The idea was that closer to the Poles, the magnetic declination would be more obvious. If this was indeed correct, this magnetic property might be employed to determine one's position on the open seas, provided that one could simultaneously determine one's latitude. This also assumed that the behaviour of the Earth's magnetic field was known sufficiently well. Magnetic declination tables could then provide the final piece of the puzzle to determine the corresponding longitude, since the direction of true North could be determined from the Sun's meridian passage (or, at night, the direction to Polaris, the North Star).

At the request of the Dutch merchant navy, on 3 September 1593 the States of Holland resolved to hear Plancius explain his method. The governing body appointed a committee of *theoristen*, composed of Joost de Menyn, the Heer (landlord) of Warmond; Jan (Johan) de Groot, Mayor of Delft; a deputy acting for Amsterdam and Willem Cornelisz Kort from the Noorder Quartiere, literally the northern quarter (the northern region of the present-day Dutch province of North Holland). A few days later, on 7 September, the States of Holland resolved that Plancius would receive generous financial compensation for his efforts, provided that the method proved successful during sea trials.

The first Dutch East India fleet departed from the port of Texel, on the North Sea, in April 1595 en route to the East Indies, where they arrived in 1596. Plancius had taught Frederik de Houtman, junior merchant of the VOC, how to measure and record compass declinations. In the mean time, he had continued to improve his method in collaboration with Mathijs Syeverts (Sieverts, Syvertsz Lakeman). In fact, in 1593 Syeverts had produced a new *portolan*-style marine map (*cf.* de Grijs, 2017: Ch. 2), equipped with additional navigation aids (such as graduated arcs). He claimed that,

> … all pilots, navigators and skippers would [be able to] sail South and North, as well as East and West, with certainty, … and [they] could know where they were without [the need for] any guessing at what distance and how far to the East, West, South or North, given the prevailing conditions. (Davidse, 2000–2020: Note 3)

Syeverts appears to have been inspired by a rather curious voyage referred to in his 1597 manuscript (Syeverts, 1597), *A very helpful treatise for all sailors/based on a discussion between two pilots*[9]. Its subtitle is telling, "Many helpful things revealed by the pilots/in particular the highly sought-after art to find East and West/and observations of the same", followed by the maxim, "Who is it who knows/other than he who fits and measures".

The treatise includes an anecdote about one Mathias Sofridus who, in December 1595, equipped his ship with wheels and wings to allow it to travel at high speed across the Arctic ice sheet. This apparently allowed him to discover a northern passage between Asia and North America, known as the mythical Strait of Anián. As the anecdote goes, during the voyage the navigator Fantanus discusses his navigation equipment with Neptune, god of the sea. He specifically refers to a particular type of magnetic compass, a quadrant and a special, three-legged 'pair of compasses'.

On 26 June 1598, upon de Houtman's return from the Dutch East Indies, Plancius and Syeverts requested that the States of Holland examine their newly improved instruments once again, in the hope of securing an additional financial reward and reimbursement of their expenses. A new assessment committee was appointed, now including intellectual heavyweights such as Scaliger, Rudolf Snellius, Stevin, van Ceulen and Waghenaer. Although there are no records as to the eventual outcome of the committee's assessment, a resolution passed on 21 May 1601 states that both men would be awarded 150 pounds if they were willing to have their method and instruments tested at sea by six to eight experienced sailors (de Jonge, 1862: 84; Davidse, 2000–2020).

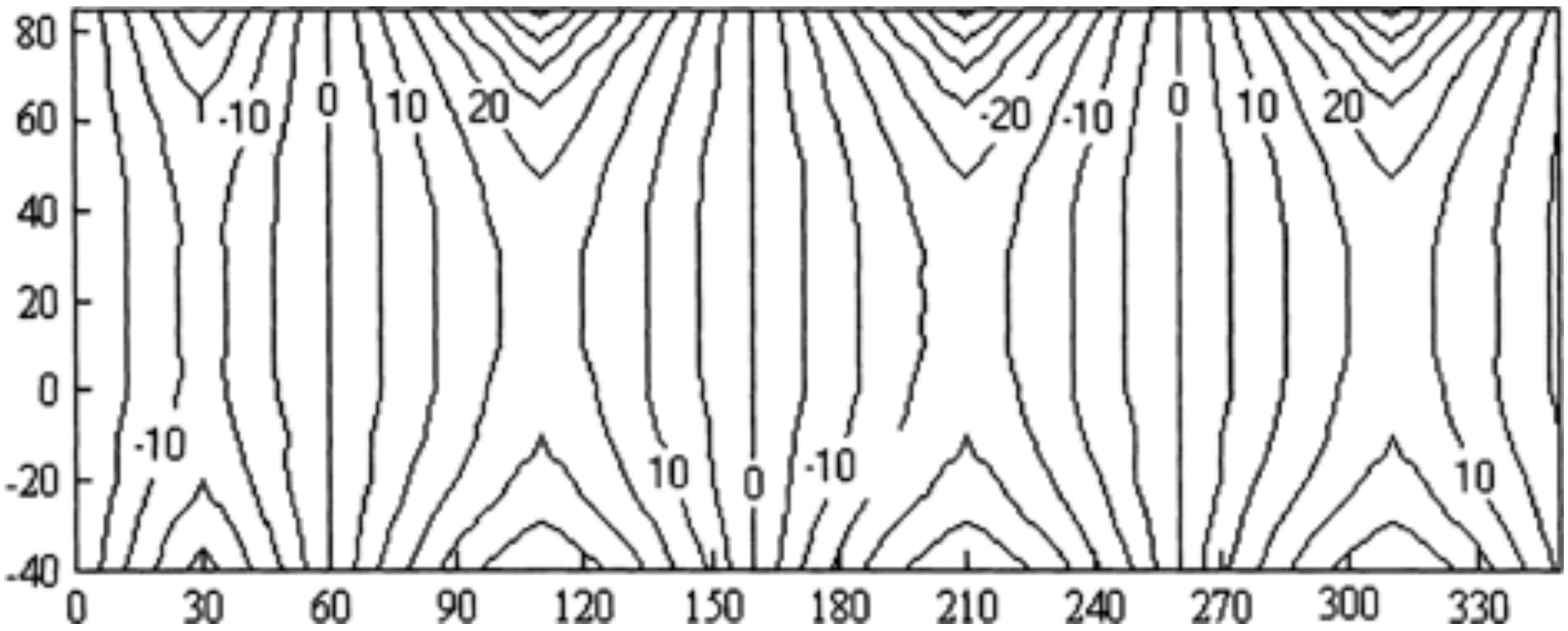

**Figure 8**: Model agonic meridians according to Plancius. (Courtesy: dbnl, Digitale Bibliotheek Nederland; not in copyright)

Plancius' theory was based on the assumption that on four meridians on Earth the magnetic declination was zero. These 'agonic' meridians (see Figure 8) included the prime meridian at Corvo (Azores) as well as reference meridians at 60° (at Cape Agulhas, South Africa), 160° (at Canton; present-day Guangzhou, China) and 260° East (at Acapulco, Mexico) (Jonkers, 2005). The areas between subsequent reference meridians are known as 'lunes'. In each of those lunes, magnetic needles would exhibit identical magnetic declination behaviour, that is, in the northern hemisphere they would point towards the Northeast in lunes I (0–60°) and III (160–260°), and towards the Northwest in lunes II (60–160°) and IV (260–360°). Travelling from West to East, the declination would increase until the middle of the region, and it would subsequently decrease. Stevin advocated for the use of six lunes, with meridians located at 0°, 60°, 160°, 180°, 240° and 340° (note that his and Plancius' lunes I and II were identical). Nevertheless, and despite this minor disagreement, Stevin was clearly in awe of Plancius' dedication in data collection,

> … listing in a table the variations that have already been observed, which the learned geographer Mr Petrus Plancius has collected by protracted labour and not without great expense from different corners of the Earth, both far and near, so that, if navigators shall find land and harbours generally in this way, as some in particular have already found them, the said Plancius may be considered one of the principal causes of this. (Stevin, 1599)

Plancius' method was, in fact, widely used from 1596 onwards (Jonkers, 2005). It was formalised mathematically in Stevin's manuscript *De Havenvinding* (*The Art of Haven-finding*: see Figure 9) in 1599, a report most likely commissioned by the nation's head of state, Lieutenant–Admiral Prince Maurits (Maurice), an enthusiastic supporter of Dutch efforts at attaining maritime dominance over the British. Stevin does not leave any doubt about his manuscript's intentions:

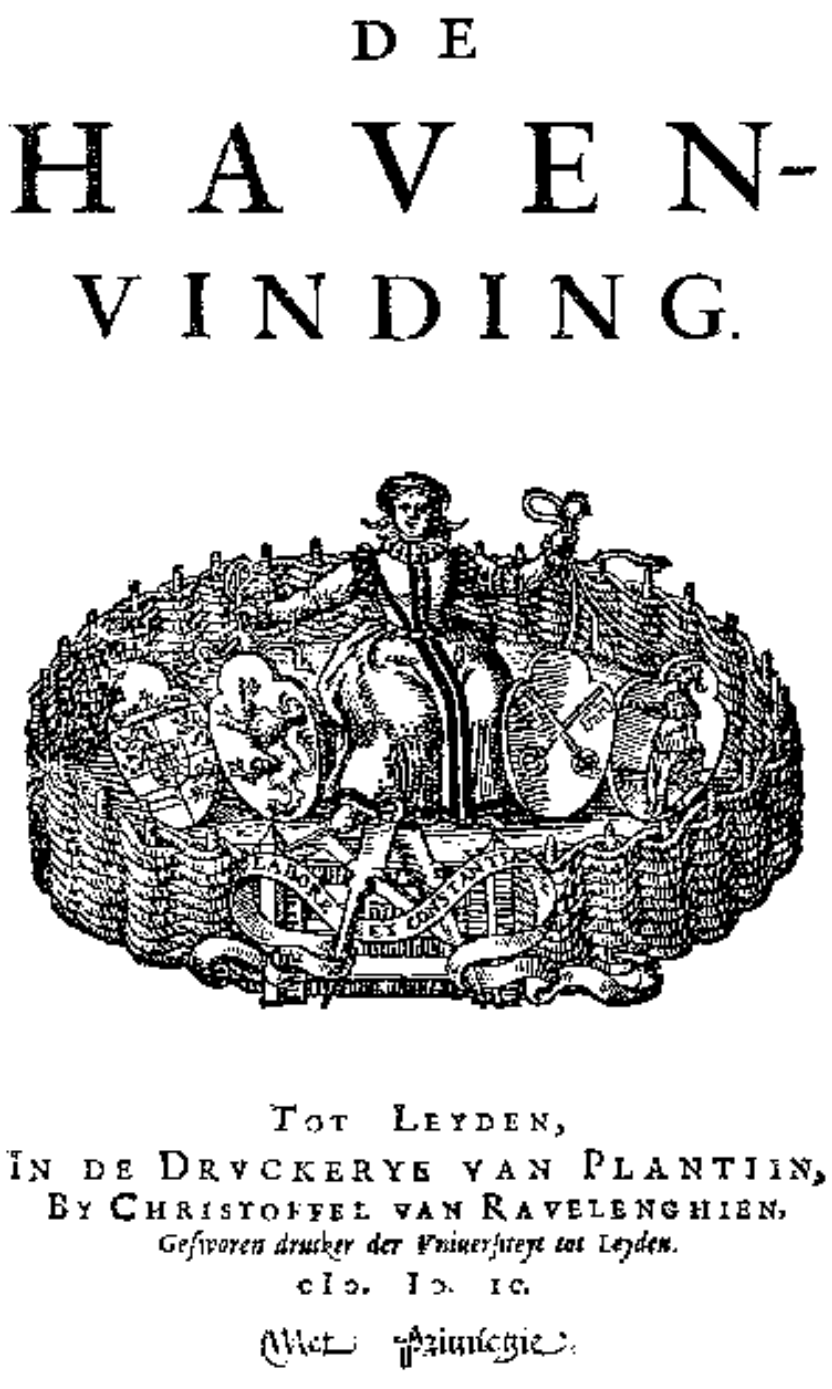
DE
HAVEN-
VINDING.

TOT LEYDEN,
IN DE DRVCKERYE VAN PLANTIJN,
BY CHRISTOFFEL VAN RAVELENGHIEN,
Gesworen drucker der Vniuersiteyt tot Leyden.
CIↃ. IↃ. IC.

**Figure 9**: *De Havenvinding*, Simon Stevin (title page). (public domain)

> It is known, that for a long time past, principally since the great voyages to the Indies and America began, a means has been sought by which the navigator might know at sea the longitude of the place where his ship is at the moment, in order thus to get to the harbours to which he wishes to go, but that hitherto it has not been possible to arrive at such accurate knowledge of the longitude. For some people, hoping to find it through the variation of the compass [the magnetic declination], ascribed a pole to the said variation, calling it magnetic pole, but it is found upon further experience that these variations do not obey a pole. Nevertheless the search for this has furnished a means for reaching a desired harbour, even though the true longitudes of both the harbour and the ship are unknown. (Stevin, 1599*a*)

Stevin's aim was indeed modest; he intended to provide a means to reach a given harbour, or even to enable ships of a particular fleet to regroup at a specific point, without knowledge of one's specific longitude:

> Since the given variation and latitude in combination indicate a definite point, both at sea and on the land, it follows from this that it is possible for ships to find each other at a given point at sea, far from the land. This is useful, among other things, to help the ships of a fleet to reassemble after a storm. By this means it is also possible to fix a rendezvous where ships coming from different directions may meet at a predetermined time. (Stevin, 1599*a*: Appendix)

A mathematician, physicist and engineer, Stevin differed from Plancius in the sense that he did not make any assumptions about the dependence of magnetic declination on geographic position. Neither was he convinced of the existence of a magnetic pole, a location usually conceived as a rocky outpost somewhere in the Arctic. Instead, he was a strong supporter of carefully executed empirical science, and so he advocated for the collection of as many measurements as possible over as wide an area as feasible to allow for a proper assessment of Plancius' theory.

His approach may well have been a forerunner to today's fashionable efforts to facilitate 'citizen science'. Stevin believed that uneducated sailors could obtain the requisite observations just as well as the average educated person (van Berkel, 1998). However, he reasoned that it was imperative to engage those sailors, and so instead of writing in French or Latin, he published his instructions and technical treatises in Dutch, using simple language. He even endeavoured to translate all remaining technical terms from Latin into Dutch for the first time. Around the turn of the seventeenth century, scholars adopted this approach more often (e.g., Stevin's contemporary Adriaan Metius also regularly published in Dutch; Dijkstra, 2012).

In his treatise *De Havenvinding*, Stevin had collected all such measurements from the *Eerste Schipvaart* to the East Indies to provide a more comprehensive and improved assessment of Plancius' method (for a less favourable opinion of Stevin's work, see Busken Huet, 1882–1884: 238, Note 1). Stevin's method of data collection was sound, which facilitated his eventual conclusion that the magnetic needle's declination was subject to secular variations. Finally, by 1611 Plancius method had actually been *in*validated based on overwhelming empirical evidence.

In retrospect, the 1590s became a turning point in the history of Dutch efforts of longitude determination. In addition to Plancius, three other inventors applied for intellectual property protection by the Dutch authorities, before a longitude prize had even been formally established: Simon van der Eycke in 1595, Reynier Pietersz van Twisch in 1597 and 1598, and—as we saw already—one Jacob van Straten in 1600.

Simon van der Eyke, also known as Duchesne, du Chesne or à Quercu (although the latter surname is likely an incorrect identification; Bierens de Haan, 1878: 7), had settled in Delft by 1584, where he taught mathematics. He is notorious for his treatise on the (incorrect) calculation of π via examination of the quadrature of a circle ('squaring the circle'), which was severely criticised by his contemporaries.

In an attempt to redeem his damaged reputation, he asked the States of Holland to issue him with a *privilege* for a new instrument he had developed to "find the East and West" (Davidse, 2000–2020), and which would allow him to produce more accurate maps. On 12 September 1595, the provincial government agreed to offer the inventor "honest compensation so that he … will be satisfied and content" (Bierens de Haan, 1878: 99), provided that his invention turned out to be practical for use at sea and that he would disclose, within three months, all technical details to a committee appointed by the governing body. However, since van der Eycke also indicated that his instrument would be useful to determine one's latitude, that is, the instrument was said to allow the user to measure "longitude both in the East and West and … latitude in the South and North", the invention was most likely simply an improved means of determining the altitudes of celestial objects (Molhuysen et al., 1911–1937).

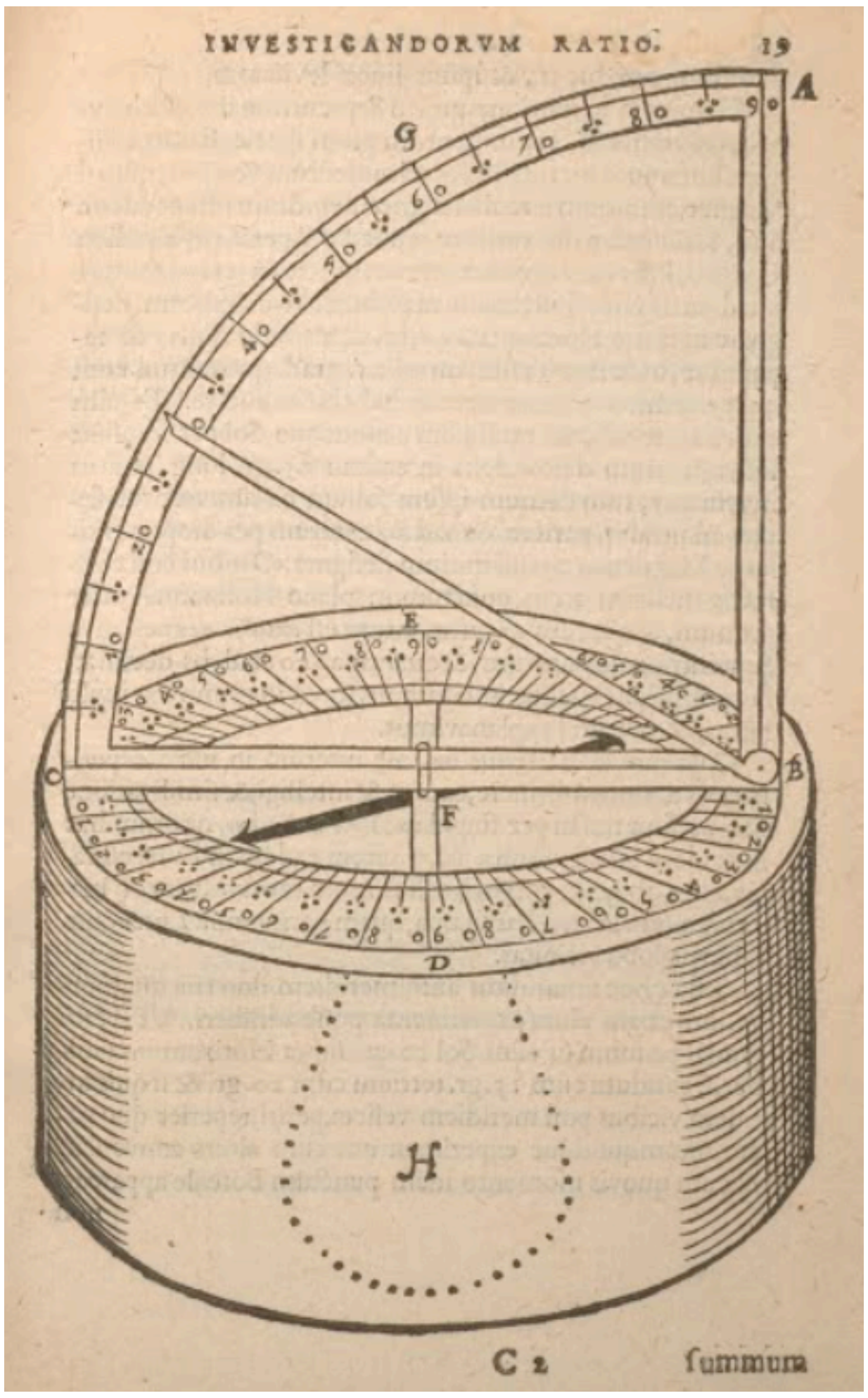

**Figure 10**: The 'golden compass' of van Twisch: *Quadrantem Azimuthalium seu verticulu cuius planu horizontale*, that is, an azimuthal quadrant that turns about a vertical axis over a horizontal graduated circle. (Stevin, 1599b; public domain)

Meanwhile, Reynier Pietersz van Twisch (Twisck, vant Wisch), on the other hand, had developed a truly novel instrument that was commonly seen as having great promise. Van Twisch, who counted Plancius among his friends or at least among his acquaintances (Busken Huet, 1882–1884: 237), was an experienced navigator with the Dutch merchant navy, and so he came with excellent credentials. On 8 March 1597, the States General awarded him a *privilege* for twelve years to construct an instrument that could easily measure the altitudes of the Sun and the stars at sea (Davidse, 2000–2020). Three days later, the commission of experts retained to assess van Twisch's proposal additionally awarded him a patent for an instrument that "by measuring the [altitude of the] Sun could allow [one] to know the deviation of the needle [the magnetic declination] as well as the longitude", an instrument that would float and which became known as the 'golden compass' (Collot d'Escury, 1829: 295).

In early March 1598 van Twisch applied to the provincial States of Holland for a subsidy to construct two instruments. One of these was a new invention known as a *Quadrantem Azimuthalium seu verticulu cuius planu horizontale* (translation into Latin: Stevin, 1599*b*: 19), an azimuthal quadrant that turned about a vertical axis over a horizontal graduated circle:

see Figure 10. Upon receipt of his application, on 13 March 1598 the States of Holland appointed a committee composed of Scaliger, Rudolf Snellius, van Ceulen, Stevin and deputies of Amsterdam, Rotterdam, Hoorn and Enkhuizen to examine the instruments and report on their performance during sea trials. Although the committee's conclusions are unknown, Stevin recommended adoption of the instrument. This suggests that the *theoristen* deemed the device practically useful.

The device was kept horizontal by floating on the surface. It consisted of a quadrant that was placed on a round compass "in the form of a double quadrant", with a sight channeling a beam of sunlight onto the compass, which was equipped with a graduated circle (Blok and Molhuysen, 1912: 1458–1459). The angular direction thus indicated by the beam of sunlight, presumably at the time of the Sun's meridian passage, would directly yield the compass needle's deviation with respect to true North. For the first time, these magnetic declinations were expressed in angular rather than temporal units, adopting common practice similar to latitude determination (Abbing, 1841: 132).

Van Twisch demonstrated the successful performance of his device on a tour around the Dutch provinces of Holland, Utrecht and Friesland. He also received ringing endorsements from the influential merchant and historian Jan Huyghen van Linschoten of Enkhuizen and from a number of experienced sailors and navigators on voyages to the African coast at Guinea, as well as the East and West Indies. In addition, Stevin included a description and drawing of the instrument in *De Havenvinding* and its Latin and French translations (see also his subsequent publication: Stevin, 1608), clearly recommending that navigators should use "an azimuthal quadrant, the horizontal plane of which, notwithstanding the movement of the ship, always remains level" (Stevin, 1599*a*).

Yet, Plancius' golden compass was severely criticised by the cartographer and experienced navigator Aelbert Hendriksz Haeyen in his manuscript, *Corte onderrichtinge* (*Brief instructions*; Amsterdam, 1599; Blok and Molhuysen, 1912). As a consequence, van Twisch declined an offer from the States General of a 5,000 guilder reward, combined with a 1,000 guilder annuity. On 3 September 1611, he once again submitted a new instrument design to the States General, this time in collaboration with Gerrit Pieters, and again the government convened a committee of experts. Despite his successful applications, in 1613 van Twisch died an old man of small fortune.

### 2.2 A cottage industry emerges

The announcement of the first Dutch longitude prize in April 1600 led to a flurry of activity, with numerous 'projectors', from genuine scientist-scholars to lunatics and those keen to cash in on the generous cash prize clamouring for the attention of the governing bodies. This is reminiscent of what happened in Spain around the same time (de Grijs, 2020*b*) and of what would happen following the establishment of the British Longitude Prize a century later (de Grijs, 2017).

Nevertheless, both the States General and the provincial States of Holland were forced to take any and all applications seriously, given the enormous economic and political ramifications associated with developing a viable solution to the longitude problem. As we will see shortly, this compelled the governing bodies to continue to engage even with those projectors whose ideas were clearly untenable.

One of the first promising applications following the establishment of the prize, once again exploiting magnetic declinations, was championed by Barent Evertsz Keteltas. On 19 August 1608, he applied to the States General for a patent and subsidy to construct a new instrument. Prince Maurits and members of the States of Holland discussed the merits of Keteltas's proposal at a meeting at the prince's quarters on 17 October that year. A decision would not be made before 22 October. However, Prince Maurits took the opportunity of the 17 October meeting to already inform the States of Holland of the conclusions of the committee of experts that had been enlisted to examine the invention (Vermij, 2010).

Keteltas proceeded to write a book to support his invention, *The use of the compass*

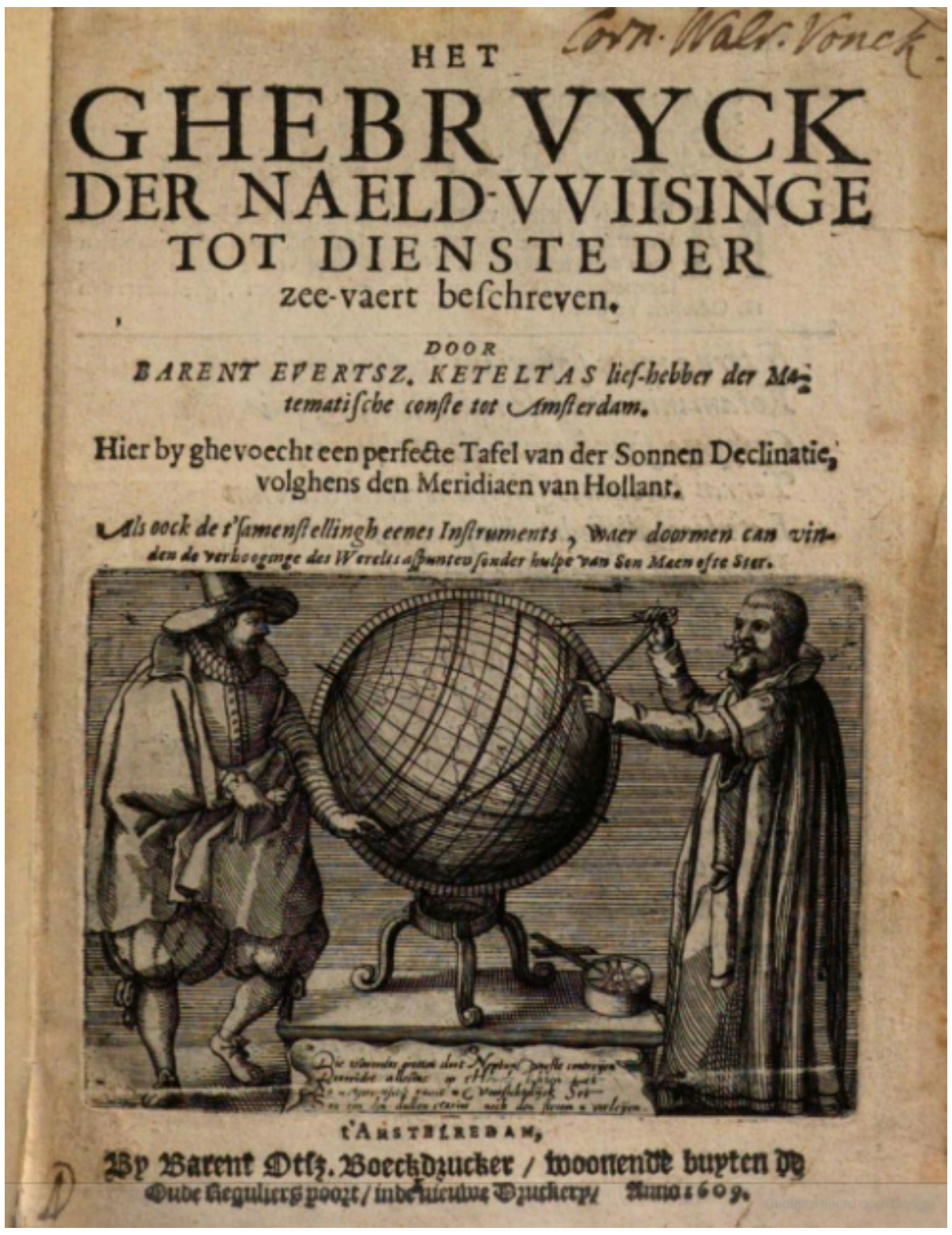

**Figure 11**: *Het Ghebruyck der naeld wiisinge tot dienste der zee-vaert beschreven* (Keteltas, 1609), title page. (public domain)

*needle in service of maritime navigation*[10] (Amsterdam, B. Ortz, 1609: see Figure 11), which he dedicated to Prince Maurits and the Admiralty of Holland, Zeeland and Westfrisia. He also presented copies to the Amsterdam City Council. In response to his advocacy, on 10 August 1610 the City Council gave him 25 guilders (Laboranter, 1862), while the States General awarded him 150 pounds (Laboranter, 1871; Smit, 1925) on 2 September 1610. However, the States General also made it clear that he should not expect any additional financial compensation for any further development of this device (Davids, 2000–2020). Although there is no surviving evidence of Keteltas's continued involvement in nautical instrument development, his name re-appears a few years later as an expert committee member retained to assess an application for the longitude prize by Jarichs van der Ley.

An excellent example of how far the Dutch government felt it needed to go in its pursuit to secure a practical longitude solution is offered by their protracted engagement with the Englishman Thomas Leamer (Leoman, Lyamor) of Amsterdam. By all accounts, Leamer was an unpleasant person in his interactions (Sprunger, 1993: 81–82) and lacked even a basic understanding of mathematics and astronomy. As early as the winter of 1609, Leamer submitted a request for a share of the Dutch government's longitude prize, at the time rumoured to be as high as 25,000 guilders (Blok and Molhuysen, 1912: 790–791). We first formally hear about Leamer from a resolution issued on 28 August 1610, when the States General awarded him a patent, although without committing to offer an assessment of the method's added value or viability (Davidse, 2000–2020).

Self-taught in Hebrew, Leamer's interest in the longitude problem was apparently inspired by insights he had gained from his religious studies, combined with a few short ocean voyages he had undertaken and his interactions with acquaintances associated with the VOC (Blok and Molhuysen, 1912). One of the key elements of his proposed new method consisted of assigning numerical values to Hebrew words so as to discover their hidden meanings. Specifically, he marked his instrument's graduated arc with letters spelling out *E-L-O-H-I-M*, the Hebrew name for god. Leamer was convinced that he had worked out a novel method of longitude determination similar to the lunar distance method, although substantially different from the practical implementation of Plancius.

Despite the initial non-committal response from the States General, the governing body issued two additional resolutions in response to Leamer's correspondence. On 24 March 1611, the States General resolved to provide Leamer with a copy of the resolution in which its longitude prize was announced. Next, on 4 July 1611, we learn that Leamer had offered a demonstration of his method if the States General would advise him on the size of the award. The government decided to send deputies representing the provinces of Holland, Zeeland and Friesland to appease the applicant and to notify him that he would be considered for a reward of 10,000 guilders. This was increased to a maximum of 15,000 guilders in a resolution of 7 July, at Leamer's request (Davidse, 2000–2020).

Rudolph Snellius and the nautical expert Robbert Robbertsz le Canu were tasked with assessing Leamer's proposal. Despite their unfavourable opinion—"more a preacher's than an astronomer's thesis" (Blok and Molhuysen, 1912: 790)—they offered him a chance at redemption by asking him to calculate a number of example problems. It transpired, however, that Leamer had not even mastered the basics of mathematics and astronomy (de Wreede, 2007), and so his 'invention' was deemed useless and futile. On 6 July 1612, the Admiralty of Amsterdam concurred and resolved to return Leamer's manuscripts. In response, Leamer proceeded to publish his ideas, a description of the method and a refutation of the objections in a confusing and lengthy pamphlet, *Clear demonstration of how to find one's meridian longitude using the timepiece of Elohim or the great Owner, namely by means of the Sun, Moon and stars at all positions in the world*[11] (Kampen, 1612). He offered his manuscript to the States General for their consideration on 12 October 1612.

Undeterred by the fact that Prince Maurits had decreed that all Dutch ships were to be equipped with instruments for longitude determination according to Plancius' design, Leamer bluntly declared that Plancius' approach was unsuitable (Blok and Molhuysen, 1912: 790–791). He appealed the rejection of his method, arguing that they were biased against him. This prompted the government to appoint a new committee, including the famous cartographer Blaeu, but once again the commission's verdict was that Leamer's proposal was "vain and frivolous" (Ploeg, 1934: 50). In addition, the protestant theologian Abraham Costerus penned a devastating response to Leamer's religion-infused pseudo-science, *The horrific, unheard-of blasphemy and rage of Thomas Leamer*[12] (1613). The dispute escalated until well into 1615, to the extent that Plancius himself was pulled into a public debate about the relationship, if any, between Hebrew letters and the art of navigation (Nauta et al., 1988: 245). The text of this debate is no longer available.

While this polemic with Leaman was playing out, the Dutch government also entertained novel and significantly more promising ideas proposed by Court (Courdt, Coert, Coenraed, Coenraedt) Boddeker (Boddecker, Borreker, Borckel) of Bremen. Boddeker had constructed a device he referred to as a *globum* (globe) or a "*cooperen granaetcogel*", a copper ball the size of a fist, on which "the longitude and latitude of the world with all of its regions, degrees and numerous indications" had been inscribed (Blok and Molhuysen, 1914: 125–126), so that "all sensible skippers and navigators will be able to find and measure the degrees, both East and West, and South and North, without fail" (Dodt van Flensburg, 1846: 261). In addition to this device, on 21 May 1612 Boddeker and Jarichs van der Ley jointly offered the States General a 24-hour hourglass, with the request that they be reimbursed for the expenses incurred (Davidse, 2000–2020).

Duly impressed, on 23 May the Dutch parliament resolved to award the pair 15,000 guilders if the invention stood up to rigorous testing at sea. In response, Boddeker offered the States General "a certain written model", while also requesting that the invention be examined by "sensible" navigators (Blok and Molhuysen, 1914: 125–126). The government subsequently decided, on 12 June, that Boddeker would be asked to construct and deliver the instruments described in his manuscript. The following year, he would also need to demonstrate the practical viability of his approach to a committee composed of members retained by the Admiralty of Amsterdam (which was informed by the States General on 18 June) and the VOC (Davidse, 2000–2020). Once again duly impressed, the committee of experts approved the invention, except for "an instrument with which he could measure [the] time" (Blok and Molhuysen, 1914: 126). This approval was followed by a formal resolution to the same effect.

Given this favourable assessment, in July 1613 Boddeker requested that his expenses be covered to the tune of 400 guilders, a request he later supplemented with a patent application. The government eventually approved both, offering him an advance on 8 January 1614. Curiously, nothing else is known about Boddeker's subsequent efforts, other than that he apparently left The Hague some time that year, 364 guilders in debt and leaving behind a "pretty copper globe that is worth much more" (Blok and Molhuysen, 1914: 126). The States General compensated his creditor on 27 November 1614, offering an advance payment of 250 guilders. Boddeker resurfaced the following year, since on 2 June 1615 he requested not only that his method be examined by the most experienced sailors and navigators, to be appointed by the Admiralty, but he also offered to take his instrument to the test during a sea trial (Davidse, 2000–2020). The States General resolved to pass on this request to the Admiralty of Amsterdam, but no further correspondence is available as to the outcome.

Also around this time, Metius' *Basic Astronomy Education*[13] (1614) appeared, in which the author—an accomplished astronomer and nautical expert—suggested that one's longitude might be obtained by observations of the mountains on the Moon, which by that time could be observed quite easily. After all, Hans Lippershey's 'spyglass' had facilitated the development of Galileo's telescope already by 1609 (Vermij, 2010). However, Metius was clearly aware that any method of time measurement remained inherently uncertain in the absence of a reliable timekeeper (Ploeg, 1934). In addition, one would need to take into account the Moon's libration, which was not well understood at the time.

Meanwhile, a few other hopefuls attempted to obtain a share of the generous prize money. First, we learn from the historical collection of resolutions passed by the States General that one Abraham de Huysse of La Rochelle must have submitted one or more instruments for consideration by the Dutch parliament (Davidse, 2000–2020). On 18 February 1616, the governing body resolved to have a demonstration of his instrument proposal assessed by Willem van Driel, Magnus and Stevin. The historical trail goes cold following that resolution, however, and we do not hear from de Huysse again.

An unusual method was suggested by Jan Jansz Stampioen, 'the Elder', teacher of navigation to skippers and first mates in Rotterdam. Stampioen was also a surveyor, an accomplished cartographer and an inspector of weights and measures. He may have been a navigator himself earlier in his career, since he is said to have sailed to the Arctic (Blok and Molhuysen, 1912: 1356–1357). From experience, he believed that sailors could navigate without the use of instruments and determine the altitude of Polaris without any problems.

On 27 July 1617, the States General received a request from Stampioen to be granted a 12-year *privilege* to teach sailors and navigators, and publish, four new methods to determine the altitude of Polaris. It appears that these methods could, in turn, be used for position determination (Davidse, 2000–2020). The government resolved that Stampioen would have to pass an examination prior to being afforded this *privilege*. Stevin and a number of deputies organised a suitable assessment on 29 July 1617, returning a positive recommendation.

Later that year, on 29 December, the Admiralty of Rotterdam enlisted a committee, which included Stampioen's colleague David Davidtsz and captain Kunst, for further assessment of Stampioen's methods. On 27 January 1618 he was offered 400 guilders as reward for his invention. The States General offered him an eight-year patent for his manuscript, *New tables to measure the altitude of Polaris, etc*.[14] (Rotterdam, 1618) and, on 10 April 1619, a financial reward of 150 guilders (van der Aa, 1874: 945; Blok and Molhuysen, 1912: 1356–1357).

## 3 DEAD RECKONING: TAKE 2

In 1616–1617, proposals from two credible applicants significantly increased the stakes of the longitude competition. Abraham Cabeljau (Cabeliau, Cabeliaeu) and Jarichs van der Ley had both been working on modified and improved versions of the traditional technique of dead reckoning, supplemented with trigonometric calculations. It is possible that both men knew each other more than cursorily, given that Jarichs van der Ley's patent application of 28 October 1617 imposed identical conditions to those requested by Cabeljau in his applications of June 1617 (Molhuysen et al., 1927: 259–260). Although the use of trigonometry soon became a standard approach in navigation, Cabeljau and Jarichs van der Ley were early adopters. Jarichs van der Ley's proposal (see below), in particular, was a forerunner of developments to come. It is no wonder, therefore, that his approach became one of the few methods that made it to actual sea trials.

Cabeljau, a bookkeeper from Amsterdam,[15] initially approached the States General in 1616 to advertise his new invention, "by means of which one could find all compass directions, both the longitude East and West and the latitude South and North" (Molhuysen et al., 1927: 259–260). In response, on 9 November 1616 the States General agreed to issue him with a patent, provided that the invention proved practically viable. The government hence ordered that sea trials be organised; a resolution issued on 16 December suggests that Cabeljau followed up by providing additional instructions and background information. On 1 June 1617, the States General rewarded him with 100 thalers for his manuscript, *Arithmetic for long-distance shipping, for determination of all compass directions both the longitude East and West and the latitude South and North, practised, described and revealed in the service of all sailors*[16] (Amsterdam, Paulus van Ravesteyn, 1617).

From a resolution of 13 June 1617 we learn that Cabeljau requested a reward—a request he repeated on 29 June of that year—and suggested that his new method be put to the test on war ships cruising in the vicinity of Cabo de Finis Terre (Cape Finisterre, northwest Spain), Tercera (Azores), the Canary Islands and Cabo de St. Vincent (Cape St. Vincent, southwest Portugal), since "there the uncertainties affecting maritime navigation are greatest" (Davidse, 2000–2020). On 14 June, his request was passed on to the Admiralty of Amsterdam, which was asked to "secretly" design a set of instructions to carry out the requisite trials. The Admiralty's recommendation, sent to the States General on 18 November 1617, was unfavourable, declaring Cabeljau's method "frivolous" and hence ineligible for a reward. As a result, on 6 December the States General resolved to reject his proposal, returning his annotated manuscript the following day (Molhuysen et al., 1927: 259–260).

Meanwhile, however, Jarichs van der Ley's star had risen, while his proposals had matured significantly from their early inception in 1612 (Historische Vereniging Noord-oost Friesland, 2011). A mathematician by training and Receiver General (tax collector) for the Frisian Admiralty in Dokkum by trade, he had filed a patent application as early as 1612 for a new '*Generale Grondregel*' (general ground rule) to potentially solve the longitude problem. Jarichs van der Ley was a keen supporter of exact, rule-based approaches, aiming to leave behind the era of sailing by intuitive estimates, a time of 'inconsistency' (Schotte, 2019: 83). Application of his method allowed for a straightforward confirmation of one's latitude using simple astronomical observations, which would simultaneously assist in correcting any errors in the estimated longitude. However, over the course of extended voyages any errors in longitude incurred anywhere along the route could still be compounded, thus limiting the method's intrinsic accuracy in mapping one's destination.

The States General retained a committee of *theoristen*, including Stevin and Samuel Marolois, mathematician and military engineer, for an initial examination of Jarichs van der Ley's method. Their initial, favourable assessment suggested that it deserved a more detailed study by experienced navigators. That jury, which included Plancius, Keteltas, Willem Jansen (most likely Willem Jansz Blaeu), Sybrand Hanssen, the navigator Hendrik Reyers and Hessel Gerritsz, eventually judged it unfavourably (Historische Vereniging Noord-oost Friesland, 2011). Jarichs van der Ley was particularly unhappy, because he perceived that the members had reached their unfavourable conclusion already prior to having assessed his approach in any detail. Therefore, he requested that the jury's comments be disclosed in written form, to which four of the committee members eventually consented.

It took Jarichs van der Ley about five months to provide a detailed written response and exposition of his method in the form of a manuscript that became a seminal work on practical maritime navigation, *The golden seal of ocean navigation, etc.*[17] (Leeuwarden, Abraham vanden Rade, 1615). He wrote this treatise to defend his approach against its critics, so it took the form of a series of dialogues combined with generally glowing profiles of his critics (Historische Vereniging Noord-oost Friesland, 2011). Despite his strong objections and the clear explanation of his method's basis, the four committee members could not be swayed. Yet, Jarichs van der Ley persisted in lobbying the Admiralty of Amsterdam, meanwhile accusing the committee members of incompetence and undeclared conflicts of interest.

Despite the experts' unfavourable assessment, the Admiralty was still interested in Jarichs van der Ley's method. Its Council reached the conclusion that conflicts of interest among the examiners could indeed have played an undue role (Historische Vereniging Noord-oost Friesland, 2011). A new committee was established, including Stevin, Jan Pietersz Dou and Melchior van den Kerckhove. The Admiralty also sponsored a sea trial to the North Atlantic in 1618. The ship, the *Bruyn-Visch*, commanded by the cartographer and pilot Joris Carolus, left the Dutch Republic on 4 June 1618 on its way to Iceland and thence along the northeast coast of Greenland and Newfoundland, returning to its homeport via the Azores on 19 November 1618. Jarichs van der Ley was convinced that the voyage had secured him his share of the longitude prize, but Carolus was less enthusiastic: although the voyage had indeed contributed to determining geographic positions more accurately than previous measurements, compounding of the longitude errors during the voyage had resulted in some not altogether insignificant discrepancies. This, of course, raises the question as to the threshold required for an improved method to warrant declaring its originator worthy of a share of the longitude prize.

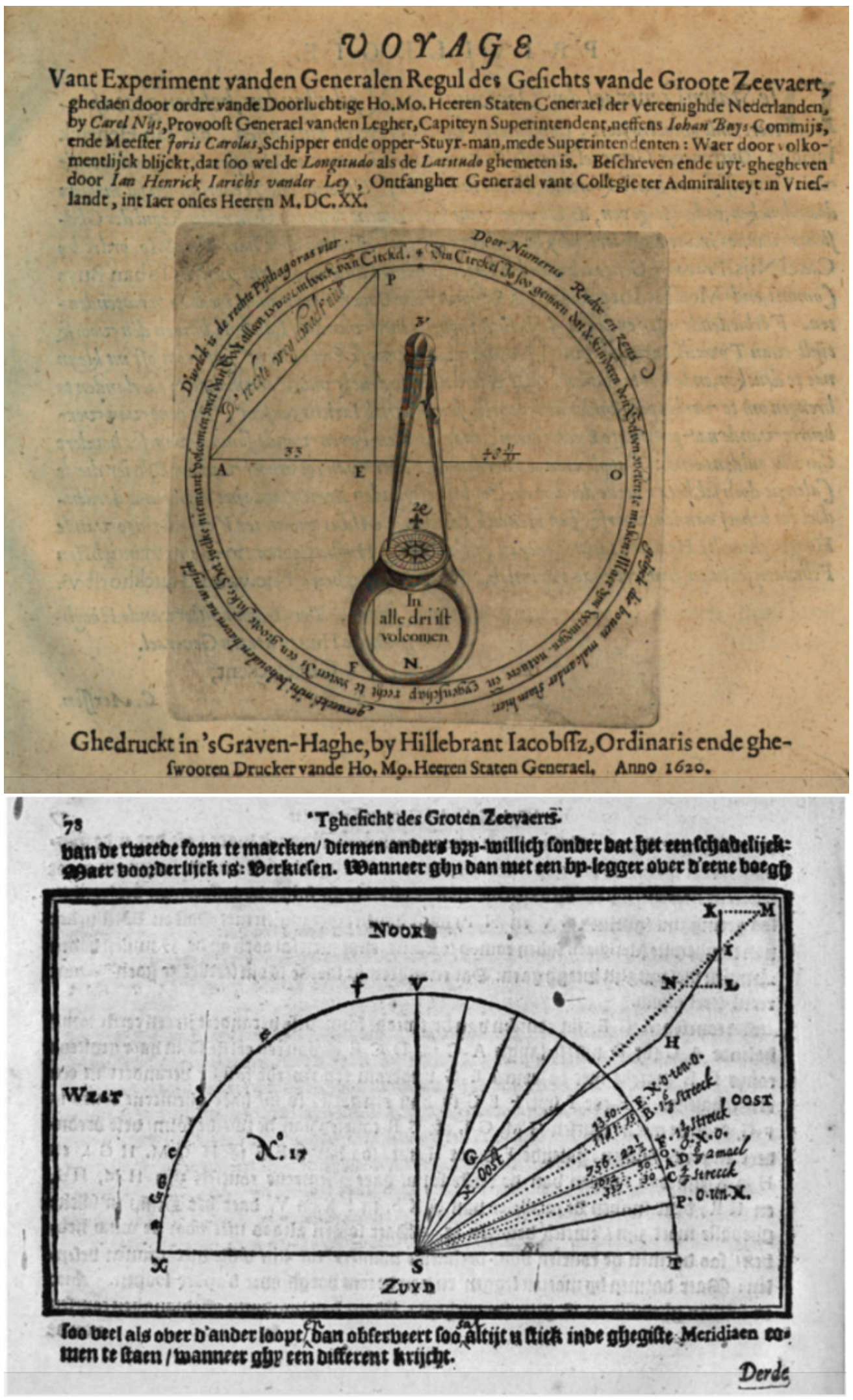

**Figure 12**: *Gesicht des grooten zeevaerts* (Jarichs van der Ley, 1619). (*top*) Title page; (*bottom*) Corrections required to determine one's longitude. (public domain)

Jarichs van der Ley nevertheless continued to pursue formal recognition of the merits of his method, which relied on transparent protractors and specially prepared paper for charting northern latitudes (Schotte, 2019: 82). In 1619, he once again published a detailed explanation of the method as well as a set of his charts in a textbook, *Gesicht des grooten zeevaerts* (*Overview of ocean navigation*; Franeker, Jan Lamrinck, 1619: see Figure 12), which was largely based on his 1615 treatise. In a cartographic context, he was among the first developers of stereographic projections, with his maps and charts showing curved parallels, meridians and loxodromes (rhumb lines that cross all meridians at the same angle) so as to facilitate more accurate position determination (see Figure 13). In fact, along with Adriaen Veen he was a pioneer in the transformation from the use of '*platte pascaerten*' (plane *portolan*-type charts) to the '*gebulte kaarten*' (spherical charts) occasionally used by the first East India fleets. Despite Jarichs van der Ley's innovations, which formed an integral aspect of the cartographic and navigation renaissance of the Low Countries, his projections were eventually not generally adopted; the Mercator projection became the workhorse approach instead.

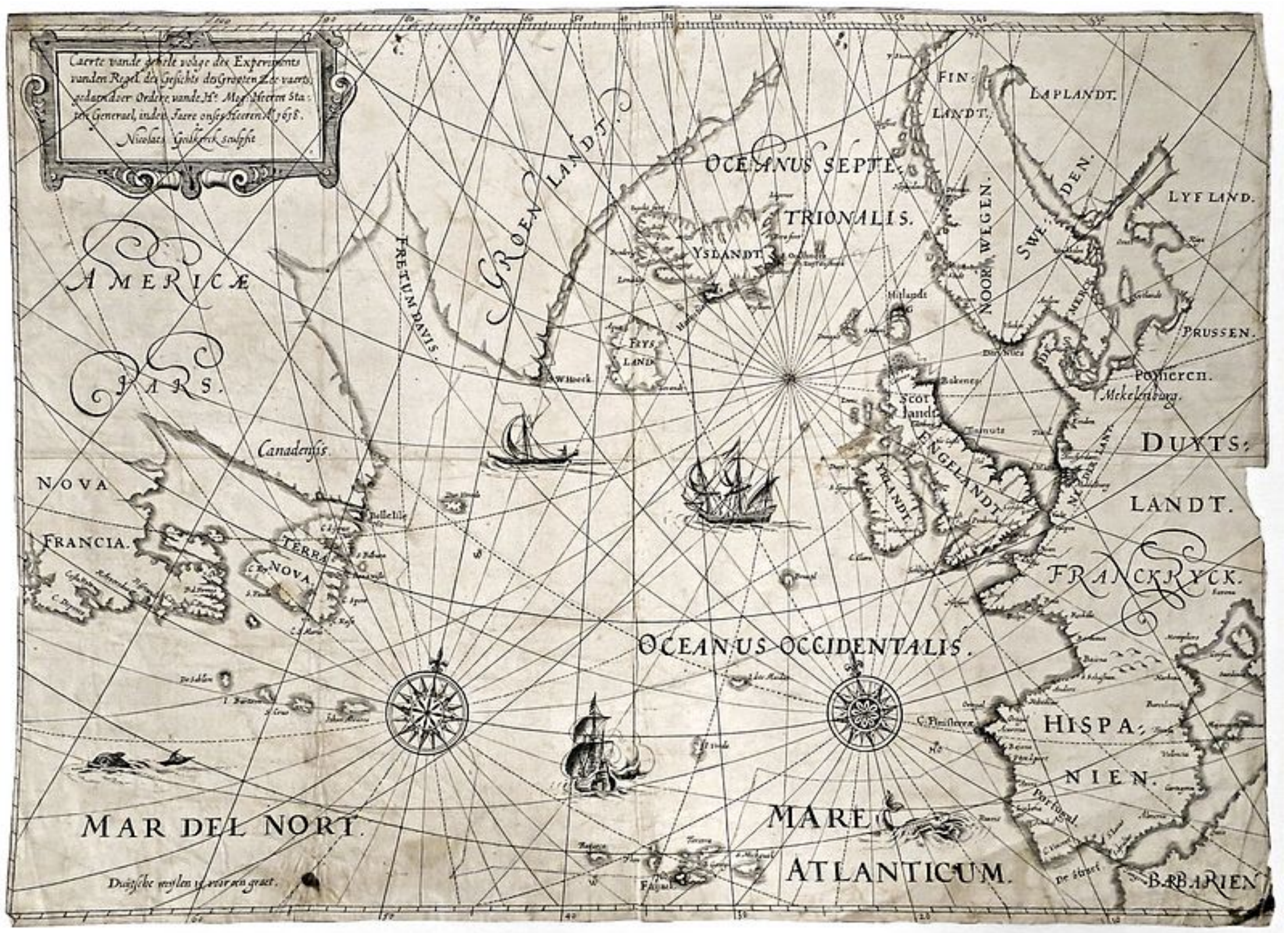

**Figure 13**: Marine chart of the North Atlantic inserted in a report entitled *Voyage van experiment van den generalen regul des gesichts van de groote zeevaert* (1620) about the attempt by Jarichs van der Ley to calculate geographic longitudes using a new type of stereographic map. (Wikimedia Commons; public domain)

Nevertheless, Jarich van der Ley's persistence eventually paid off, however, at least to some extent. Initially, the committee of *theoristen* that had been convened to examine the outcome of the *Bruyn-Visch*'s voyage once again returned an unfavourable assessment by the end of their three-year tenure, on 5 January 1619. However, a triumvirate that included Metius and Willebrord Snellius overturned that assessment the following year (de Wreede, 2007), confirming

> … that the rule to find the longitude at sea, by several *theoristen* as well as *practisijns*, and also through navigation to distant places, by means of ships of the Royal Admiralty in Holland, had been subject to experimentation and was approved, is deemed very helpful for cartographic reform and correction and in general service to navigation on the high seas of the oceans, the Lords of the States General have sent, in this year [1620], based on the report received from the *practisijns* as well as *theoristen*, letters of credence, and instructions to all Colleges, directors, ships' captains, navigators, that they will learn and know these rules forever, follow and use them, according to the letter referred to dated 21 July and signed by M. van Lyclama [Marcus Lycklama a Nijeholt], *vidit* [witnessed]. (Winsemius, 1622; Historische Vereniging Noord-oost Friesland, 2011)

On 22 February 1620, Jarichs van der Ley published a new manuscript, *Voyage to test the general rule of ocean navigation*[18] (The Hague, Hillebrant Jacobsz van Wouw, 1620). Gradually, his method continued to attract more vocal supporters,[19] yet despite this delayed vindication and the growing external approval, the method never came into general use. This was most likely owing to the continued systematic sabotage and obstruction of Jarichs van der Ley's proposals by examiners including Cornelis Jansz Lastman (van Berkel, 1998; de Groot and de Groot, 2002), instructor of navigation at the Amsterdam chamber of the VOC. In 1619, Lastman was appointed as pilot examiner and, subsequently, as teacher of nautical science to the navigators of both the VOC and the Dutch West India Company. With tacit approval of the East and West India Company directors, and facilitated by his role as member of one of the assessment committees, Lastman prevented general adoption of Jarichs van der Ley's method for more than a decade. It took until 1631 before Jarichs van der Ley's approach was taken more seriously (van Berkel, 1998). From that point onwards, his method soon became almost the only method of longitude determination in regular practical use among VOC navigators.

## 4 GALILEO'S SECOND CHANCE

Despite the significant prize money on offer, Galileo initially ignored the Dutch opportunities in favour of pursuing a share of the Spanish longitude prize (see, e.g., de Grijs 2020*b*). As early as 25 October 1627, he had received a letter from the diplomat Alfonso Antonini, his correspondent in The Hague, informing him about the Dutch prize, however:

> I have learnt that the VOC and the States [General] have announced and secured a large sum of money (they say it is around 30,000 *scudi*), to reward those who can teach how to find the longitude for use of navigation ... If I get access to the details [of your method], I promise to write to them and inform them of your work. ... If they wish to adopt the approach, which seems beautiful and great to me, I will enjoy not only having made the proposition, but having made myself useful to conclude the business successfully and forthright. And if, by chance, you wish the negotiations to be done in secret, you can count on my discretion, which has never failed and will never fail ... (Buyse, 2017: 21–22)

Galileo's initial rebuffing of the Dutch longitude competition was significant, given that the prize money Antonini had informed him about far outweighed his meagre annual professorial salary at the University of Pisa of just 60 *scudi*. Nevertheless, he was intent on securing the Spanish longitude prize instead. Although it had become clear already by 1618 that his proposal to use the ephemerides of Jupiter's moons as an accurate, natural clock was not considered favourably by the Spanish Crown, he persisted in submitting updated proposals in 1620, 1629 and 1631, all to no avail (de Grijs, 2020*b*). Meanwhile, much of his attention was also focused on completion of his last major work, the *Discourses and Mathematical Demonstrations Relating to Two New Sciences*,[20] which was eventually published in 1638 (Leiden, Elsevier).

Galileo's attitude changed after the Catholic Church censured him in 1633 for his outspoken pro-Copernican views. He needed a safer, less controversial subject to focus on, although he still pursued a solution to the longitude problem based on observations of Jupiter's moons—potentially contentious in the eyes of the Inquisition. Given his predicament, some of Galileo's Dutch friends were hoping to extract him from Italy to provide him with a safe haven in the more tolerant Dutch Republic. These efforts were led by Hugo de Groot (Grotius), former leader of the Dutch Remonstrants and a diplomat in his own right, who lived in exile in Paris and represented the Swedish government there (Ploeg, 1934).

In July 1635, de Groot informally announced to his friends—including van den Hove, professor 'in the Copernican theory' at the Amsterdam *Athenaeum Illustre*, Blaeu and Laurens Reael, former VOC governor-general and former Dutch navy admiral—that Galileo was reasonably confident to have found a way to determine longitude at sea. The Italian scholar's idea of using a pendulum clock in combination with the eclipses of Jupiter's satellites was first proposed in a letter to the leading French mathematician and geostatistician Jean de Beaugrand of 11 November 1635:

> I have such a time measurer [*numeratore del tempo*] that if four or six examples of this instrument were constructed, and if they were allowed to operate at the same time, we would find that in confirmation of their accuracy, the times measured and indicated by these timekeepers would show differences of only one second, not only from hour to hour, but from day to day and from month to month, so uniform would be their operation; these clocks are really admirable for the observers of motion and celestial phenomenon, and in addition, their

> construction is very simple and far less subject to outside influences than are other instruments which have been invented for a similar purpose. (Buyse, 2017: 27)

The ‘time measurer’ Galileo had in mind was likely a forerunner to the pendulum clock (Galilei, 1639). The pendulum clock was first developed later in the seventeenth century, partly on the basis of Galileo’s descriptions (e.g., de Grijs, 2017): “More likely they were a form of vibration counter, consisting of a pendulum bob suspended on a string which was given impulse manually or by clockwork” (Bedini, 1991: 18). In this sense, Galileo’s proposal as supplied to the States General was substantially different from his earlier proposals submitted to the Spanish authorities. Those had not included a description of a stable, working marine timepiece. Note that Galileo’s ‘time measurer’ was not meant to transport ‘standard time’ from the homeport across the ocean (as had been suggested by Gemma Frisius a century earlier; de Grijs, 2017; 2020*a*). He intended his clock to show and keep *local* time for much shorter periods between successive astronomical events, in combination with astronomical methods.

Galileo’s proposed approach was based on the use astronomical ephemeris tables, including tables of the eclipses and occultations of Jupiter’s moons, as standard time reference. He did not realise that transporting stable clock time would be the eventual solution to the longitude problem. Galileo’s proposal as eventually submitted to the States General highlighted the high quality of his telescopes compared to that of the spyglasses available at the time in the Dutch Republic, and the accuracy of his ephemeris tables of Jupiter's satellites.

Although Galileo’s advanced age prevented him from travelling to the Dutch Republic, de Groot had already enlisted van den Hove and Élie Diodati—Galileo’s friend and representative in Paris—in persuading the Italian scholar to formally present his discovery to the States General (de Waard, 1939–1953: 236–237; Garcia, 2004). The strongest argument that likely persuaded Galileo to proceed was that the committees of experts routinely appointed by the governing body were composed of first-rate scientists who would value the merits of his invention.

On 15 August 1636, Galileo (1636b) penned an elaborate description of his method for dissemination to the States General (de Waard, 1939–1953: (4) 241–244; Drake, 1978: 374), written in Italian, which he sent to Diodati in Paris. On 20 September 1636, Diodati passed on Galileo’s missive to Reael (Galilei, 1636a; 1655: (XVI) 491–492), the Italian scholar’s most important backer in the Dutch Republic. He simultaneously informed the admiral that, jointly with de Groot, he would act as intermediary in the negotiations with the Dutch government. On 23 September, Diodati also informed Galileo that he would do everything in his power to conclude the negotiations successfully (Galilei, 1655: (XVI) 489–491).

In his lengthy letter, Galileo claimed to have constructed highly accurate timepieces, whose “construction is very simple and far less subject to outside influences than are other instruments which have been invented for a similar purpose” (Galilei, 1655: (XVII) 463–469). Nevertheless, he also pointed out that his clock could only record the time from midday onwards, and that the proposed observations of Jupiter’s moons would require a stable observation platform. In fact, one could not risk losing sight of Jupiter for any length of time during the observations, since the eclipse ingress and egress timings of its moons would only last on the order of a minute.

Despite these practical difficulties, he felt that he was now well placed to calculate and tabulate these eclipses with reasonable accuracy, thus making them useful for navigation purposes. Simultaneously, Galileo wrote to de Groot and van den Hove, via Diodati, impressing upon them the need to make progress towards practical implementation of his proposal given his own deteriorating health. De Groot (1636) responded on 20 September 1636, stating that he fully supported the “most exquisite” discovery Galileo proposed, which he hoped would be useful “for all mankind”.

Once Galileo's letter had been translated by Reael and passed on to The Hague, where it arrived on 11 November 1636, a committee chaired by van den Hove and including Blaeu and Reael was appointed (de Waard, 1939–1953: (4) 245–251). Initially, they had also intended to appoint Jacobus van Gool, an Orientalist and mathematician at Leiden University (de Waard, 1939–1953: (4) 253), but instead the committee agreed to invite Beeckman to join their ranks. (In 1631 Beeckman had proposed the use of Jupiter's satellites as a celestial clock, independently of Galileo; de Waard, 1939–1953: (3) 229–230.) In their report of 5 December 1636 the States General confirmed that,

> … through his zealous research Galileo believes to have found a certain method to determine at every moment and in every place of the world, at sea as on land, the true longitude of the location, and how much more to the East or to the West this location is situated from the Meridian of any city or port, that may be chosen freely; presenting this invention with regard to the laudable reputation [to be gained for himself] and the government in this country, and also [with regard to] the premium offered to the first author who would show and dedicate [his invention] to Her Great Power. (Buyse, 2017: 22)

We learn from Diodati's letter of 16 March 1637 to van den Hove (Galilei, 1655: (XVII) 43) that by that time Galileo had not yet received any official acknowledgement from the States General, a complaint Diodati would repeat several times in his correspondence with Dutch representatives over the next few months. Four days latter, Diodati also wrote to Constantijn Huygens—Christiaan's father and secretary to *stadholder* (regent) Prince Frederik Hendrik—asking him to intervene. The elder Huygens responded positively on 23 April (Galilei, 1655: (XVII) 59–60), in French, although he emphasised that there were two important obstacles preventing Galileo's proposal from returning a favourable assessment, including the general lack of sufficiently high-quality telescopes and the instability of ships at sea.

Meanwhile, on 7 April 1637 the committee of *theoristen* presented their recommendation to the States General. The governing body resolved that a second committee, composed of commissioner Arnold van Rantwijck, Mayor Johan van Weede of Utrecht and Mayor Wolter Schoneburch of Groningen, in addition to Reael, was required to advise on the next steps. On 25 April 1637 the States General offered Galileo a gold chain with a medal valued at 500 guilders (de Waard, 1939–1953: (4) 267; equivalent to about 1,000 *scudi*) as a show of their goodwill and in anticipation of a positive outcome of Galileo's submission. However, Galileo never received the 30,000 *scudi* he had been promised for the delivery of a viable solution to the longitude problem. The States General also ordered the VOC to provide Reael with an expense account to the tune of 1,000 guilders. Nevertheless, the Catholic Pope Urban VIII prohibited Galileo to deal with the Dutch Protestant government and hence Galileo was forced to decline the reward. In addition, Reael did not pursue construction of Galileo's proposed instruments, for which the expense account had ultimately been established.

In his 23 April 1637 response to Diodati, Constantijn Huygens expressed his desire that the "great man" (Galileo) would live to see the successful development of a stable timepiece that could be used for navigation at sea. In fact, in a letter from March 1637, Galileo had already pleaded with Diodati to convince Huygens that he had come up with a solution to the stability problem, which Galileo himself forwarded to Reael in June 1637 (Galilei, 1892–1909: (XVII) 46–49 and 96–105):

> It would be a waste of time to occupy Your Lordship's attention any longer with the details. You can command artists of the utmost skill in the manufacture of clocks and other excellent mechanisms. They have only to know that the pendulum gives vibrations of exactly equal duration, whether the arc be great or small, in order to devise methods of construction of greater precision than any that I could devise. (Robertson, 1931: 86)

Galileo's solution involved a marine chair composed of a large universal joint, with one hemispherical component moving inside a second, which in turn was to be fixed to the ship. Both components had to be separated by water or oil and it was crucial to retain a gap between them, which could be achieved with eight to ten springs, all in all a rather cumbersome contraption (Ariotti, 1972; Galilei, 1892–1909: (XVI) 96–99; see also de Grijs, 2020*b*: Figure 14). Reael called Galileo's stabilising platform solution impractical, and he thought it implausible that his sailors could operate such a complex device, since they were

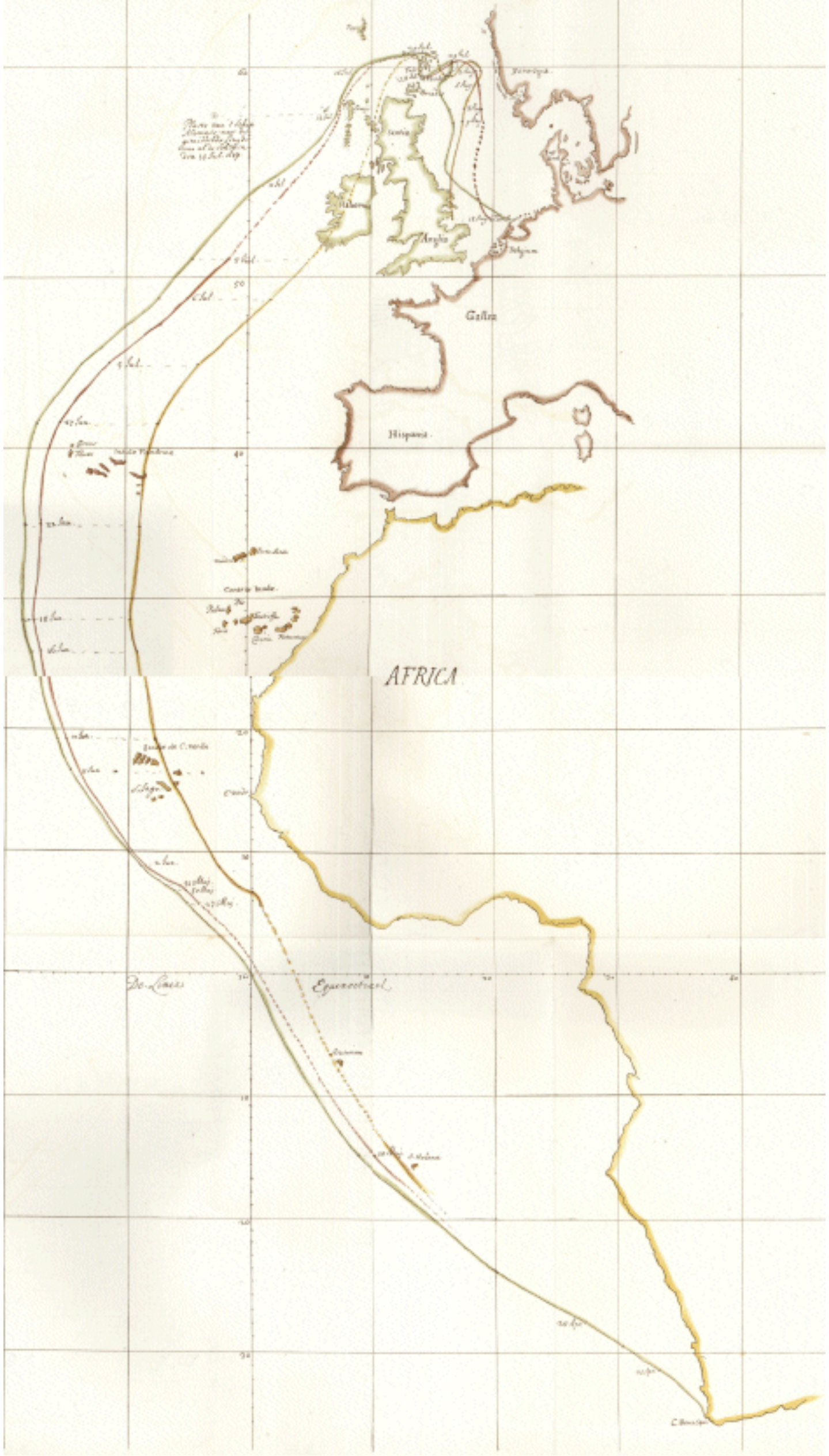

**Figure 14**: The route determined by the *Alcmaer*'s navigators is indicated by the westernmost curve (green); that based on raw timing measurements is represented by the easternmost curve (yellow). Upon correction for the effects of the Earth's rotation, Huygens arrived at the curve closest to but slightly to the east of the westernmost route (red). (*Oeuvres Complètes de Christiaan Huygens*, IX, 273; not in copyright)

> … rude people, men only superficially acquainted with mathematics and astronomy … and who still find insuperable the problem of using your discovery on a moving ship, continually being tossed about. (Galilei, 1892–1909: (XVII) 116–117; Ariotti, 1972: 368)

Diodati, meanwhile, had become impatient by the delayed response from the States General, while he was also less than pleased that Galileo's insights had been leaked, through correspondence between Beeckman and Marin Mersenne, the French scientist-priest, to one of the Italian scholar's main competitors, the French astronomer and mathematician Jean-Baptiste Morin de Villefranche (Morinus) (de Waard, 1939–1953: (4) 261–262). However,

Constantijn Huygens explained that the committee had to assess both the theoretical merits of Galileo's method and its practical application, alongside their other commitments, which implied that the process would take quite some time (de Waard, 1939–1953: (4) 267–268).

Meanwhile, van den Hove sent Diodati a lengthy letter on 27 April 1637, reassuring him that Morin de Villefranche had not been provided with any information crucial to enable him to beat Galileo to the prize (de Waard, 1939–1953: (4) 270). On 22 May 1637, Diodati responded with an invitation to van den Hove to meet Galileo in the Dutch Republic's embassy in Venice, where Galileo would be prepared to reveal additional details of this method that would, in turn, facilitate its practical viability. That meeting never took place. Although van den Hove eventually secured a government grant of 2,000 guilders to undertake the journey to Italy, via Paris (Ploeg, 1934), van den Hove's death intervened.

In addition, the enterprise which had initially looked very promising soon turned sour. The official letter from Reael containing the initial response of the States General did not reach Galileo until 23 June 1637 (de Waard, 1939–1953: (4) 258–259). His detailed response with instructions regarding the practical operation of his proposed instruments, returned to Reael on 22 August 1637, arrived on Reael's deathbed and remained unopened on his desk (Galilei, 1892–1909: 96–105, 174–175): Councillor Nicolaes van Reigersberch notified de Groot, his brother-in-law, in a letter of 25 October 1637 that Reael had passed away on 10 October:

> The loss of two children through the contagious sickness [the plague] had plunged this good man into such profound melancholy, that he forgot all other thoughts and even those that were very dear to his heart. Even a letter from Galileo Galilei, which was given to him when he was still healthy, remained unopened, which I mention so that the said Galileo may be informed of it. (de Waard, 1939–1953: (4) 282; No. 1)

In his response to Constantijn Huygens, dated 28 February 1640 (Worp, 1892–1894), Diodati (1640) referred to these and other unfortunate events that had happened in the mean time. He said that Galilei's proposal, as submitted to the States General had been affected by several interruptions. These included the complete loss of Galileo's eye-sight during the past two years (probably because of unprotected observations of the Sun; Ploeg, 1934) and the untimely death on 17 August 1639 of van den Hove, who had until recently been the sole survivor of the four commissioners originally tasked with researching Galileo's proposal. Beeckman and Blaeu had died earlier, in May 1637 and October 1638, respectively.

Nevertheless, Diodati confirmed that Galileo was keen to continue his pursuit of the construction of a timing device to determine longitude at sea. In his letter to Diodati of 30 December 1639 (Galilei, 1655: (XVIII) 132–133), Galileo proposed to send one of his students, Vincenzo Renieri, to the Dutch Republic to provide more technical details. He also suggested that the States General appoint new committee members. In fact, on 28 February 1640 Diodati asked Constantijn Huygens for assistance, given his expertise in the subject matter. The States General briefly considered convening a new committee, but that committee did not materialise (Mersenne, 1636). Despite Diodati's impassioned plea to the elder Huygens, I have not uncovered any evidence of the latter's involvement since receiving Diodati's letter from 1640, beyond a suggestion to involve Willem Boreel, VOC governor (Ploeg, 1934: 53). And so this is where Galileo's pursuits ended, unsuccessfully; he would not engage with the States General or his Dutch supporters again prior to his death on 8 January 1642.

Ultimately, the States General cited as their main reason for rejection of Galileo's proposal his method's practical unviability for use on pitching and rolling ships at sea. In addition, the governing body remained unconvinced of the reliability and accuracy of the eclipse and occultation tables (Huygens, 1662–1663; Adam, 1910), despite Galileo's assurances. Nevertheless, Galileo's method did not disappear from consideration completely. In 1661, Dirck Rembrandtsz van Nierop, the shoemaker-turned-mathematician from Nieuwe Niedorp, wrote about longitude determination at sea based on the ephemerides of Jupiter's satellites. Unfortunately, he did not manage to make the method work successfully on board a moving ship either.

## 5 TIME AND TIME AGAIN

For the next significant submission of a solution to the longitude problem to the States General, we have to wait until 1655. At the young age of 26, the States General retained Christiaan Huygens to assess a proposed new method by Jan Kołaczek of Leszno (Johannes Placentinus; Omodeo, 2016), professor of mathematics from Francfort (Frankfurt) an der Oder, for the purposes of issuing a possible patent. It is possible that his father had recommended the young Huygens to the States General, given that Huygens Jr. was invited to contribute as "the son of the Lord Zuylichem". The young Huygens was tasked with a careful assessment of Placentinus' invention. It was said to allow the "determination of East and West" using observations of the Moon, …

> … provid[ing] a way to find the longitude of places, both on land and at sea, at any time, day or night, and in this way, given the latitude of the location and having found its longitude, to determine the position of a ship as it is being swayed by storm and wanders back and forth, etc. (Placentinus, 1655)

Therefore, in March 1655 Huygens wrote to Andreas Colvius, a personal friend of Beeckman, "I expect, in turn, that you will send me manuscripts regarding the determination of longitude and whatever else you own from Galilei's legacy" (Huygens, 1655*c*). In this instance, Galilei's legacy included correspondence with Diodati, de Groot, van den Hove, Reael, Alphonse (Alfonso) Pollot(to)—an Italian officer at the Dutch government with whom Galileo had exchanged letters about the determination of longitude around 1637—and the elder Huygens (Galilei, 1892–1909: 123–191, specifically letters from 1718). In 1622 Colvius had used a journey to Venice to meet Italian scholars and collect and copy numerous books and manuscripts. At Huygens' request, Colvius (1655) sent the young scholar his notes, as well as a manuscript written by Galileo, attached to a letter dated 23 March 1655.

Placentinus had tabulated the maximum lunar elevation as well as those of the constellations Leo (the Lion; specifically for the lion's tail) and Lyra, calculated for the Frankfurt an der Oder meridian from April through June 1655, in addition to step-by-step instructions to determine one's longitude:

> First, measure during the months of April, May, and June of 1655 at your location the time at which the Moon reaches its highest elevation, either during the day as derived from the solar elevation or at night by determining star heights; such observations are not unknown to those who practice math, and also to sailors they are hardly a secret.
>
> 2°. Compare this time, observed at your meridian, with the time indicated in the table, and note the difference in the meridians between your location and that of Frankfurt [an der Oder], in hours and minutes.
>
> 3°. Convert the observed meridian difference into degrees and minutes away from the Equator, according to the second table, and you find the difference in longitude between Frankfurt [an der Oder] and your location.

Huygens' (1655*b*) assessment of Placentinus' invention was devastating. He referred to the proposed method as "Placentinus' nonsense", pointing out that the method violated basic astronomical principles, including the assumption that the Moon would traverse 15 degrees on the sky in an hour, just like the Sun. In reality, the Moon travels approximately half a degree less per hour than the Sun. Huygens stated that Placentinus' proposed method had

> … by no means as good a basis as those of others before him, who have tried to achieve the same. Their [the others'] inventions, although they were considered of little or no use (because of major miscalculations that could arise from the smallest observational errors or imperfections in their Ephemeris tables), however, were theoretically well founded. But this discovery of Johannes Placentinus is so far removed from offering any benefit or use, that it even sins against the foundations of astronomy, and it is nothing but a gross fallacy. (Huygens, 1655*b*)

Huygens' unfavourable assessment was not universally welcomed, however. The German–British polymath Samuel Hartli(e)b, for instance, was rather impressed by Placentinus' ideas:

> Placentinus Bohemus Professor Mathematicus upon the Oder at Frankford hath published a Booke De Longitudine[21] which is thought the best that ever hath been written and should not have beene thus plainly discovered the States of Holland and others having set so great a price and reward for it, which belike this Professor was ignorant of. (*Hartlib Papers*, 1655)

Surprisingly, Hartlib suggests here that Placentinus was not aware of the Dutch longitude prize, but that statement appears incorrect. After all, Placentinus applied for a patent from the States General, introducing his ideas as "[n]ew and careful research of the longitudes of places, by Dutchmen, French, English and Spanish most desired."

Following Huygens' unfavourable assessment, the States General (1655) decided to consult a number of other experts for a second opinion, including van Schooten. Huygens and van Schooten subsequently exchanged numerous letters between March 1655 and November 1656; van Schooten hence became Huygens' *de facto* mentor during the early stages of his career. Their correspondence included a copy of the *Opere di Galileo Galilei* (*Works of Galileo Galilei*). This is important as regards the Dutch efforts to determine longitude at sea, because Galileo's manuscript included ideas about pendulums.

It is possible that this first official assignment by the States General triggered the young Huygens' interest in pursuing a solution to the longitude problem. At the very least, the first notes in his handwriting on this topic originate from early 1655, when he copied a passage from Metius' 1621 treatise, *Institutiones Astronomicae Geographicae* (Ch. 4):

> ***Brief instructions as to the use of clocks to find Eastern and Western Longitudes.***
> However, this method is probably the easiest and most suitable that one could encounter (namely using clocks to find the eastern and western longitudes), its only difficulty and problem are related to incorrectly and irregularly running clocks. You, therefore, zealous researchers versed in the examination of natural things, pay attention to this, work to correct this problem and to establish the true and invariable course of nature: having established it, you will have found the true philosopher's stone and the brave sailors will no longer run into danger so often. (Huygens, 1655*a*)

Huygens' own pursuit of a practical solution to the longitude problem is well-known and has been covered by many authors, including in detail in my recent monograph, *Time and Time Again: Determination of Longitude at Sea in the 17th Century* (de Grijs, 2017). For an in-depth discussion of Huygens' place among the thought leaders of his time, I refer the interested reader to the latter publication. In the remainder of this section, I will summarise his achievements as they pertain to the Dutch longitude competition and his interactions with the Dutch government in pursuit of the prize.

Irrespective of the numerous contributions by other scholars, Huygens is rightly known as the first person to combine the pendulum—a metal ball suspended by a silk thread—with the clock mechanism. (However, note that Richard Harris of London is credited to have converted a church clock to employ a pendulum in 1642; Reid, 1826: 179.) On 12 January 1657, he wrote to van Schooten that

> … one of these days, I invented a new type of construction for a time piece, which can be used to measure times so accurately that there is more than a little hope that this can be used to determine the longitude, at least as regards travel on the seas. (Huygens, 1657*a*)

Huygens was, at least initially, keen to discuss his new invention with van Schooten and his peers. He proudly announced his invention to the French mathematician Claude Mylon on 1 February 1657, although without providing any detail:

> The news that you tell me regarding the journey of Mr. Bulliaut [Ismaël Boulliau; French astronomer and mathematician] to these lands rejoices me very much … I also want to show him one of my new inventions, which will be of the greatest benefit to astronomy, and I

> sincerely hope to apply it successfully in search of longitude. You may hear more about in a little while. (Huygens, 1657b)

It appears that by this time the Dutch longitude competition had run its course, however: in 1658–1660 the English polymath Robert Hooke explored the possibility that his own novel, spring-driven watches might be eligible for any of the awards issued for “finding the longitude”. However, the legendary rewards appeared to no longer exist on the European continent (Robertson, 1931: 172).

Nevertheless, in summer 1662 Huygens wrote several enthusiastic letters to both his brother Lodewijk and to Sir Robert Moray, the Scottish statesman, referring to a

> … small pendulum clock ... which works sufficiently well to serve for [the purposes of] longitude determination, and which, once I have given it a push, continues to move without stopping in my room, where it is suspended from 5 foot long ropes, but I have yet to test it on water, for which we should requisition a reasonably sized vessel to [allow us to] sail on choppy seas, something I do not know when I could achieve it. (Huygens, 1662)

The provincial States of Holland (1664) eventually issued a fifteen-year patent for his marine clock on 16 December 1664. Meanwhile, Huygens had realised that he could potentially make a profit from his work, particularly now that his first marine clock looked like it might sustain rigorous sea trials. Therefore, he temporarily halted his plans to publish his new treatise, *Horologium Oscillatorium*, which contained full technical details and which he had been working on since making the first revisions to its precursor, *Horologium*, in 1660.

By 1682, Huygens’ sustained efforts to perfect his marine clock design had become sufficiently promising to attract the attention of the governors of the VOC. As a result, the VOC issued a first resolution of support on 31 December 1682 (VOC governors, 1682). It specifically authorised the Mayor of Amsterdam, Johannes Hudde, governor of the VOC and a mathematician by training, to …

> … correspond with Mr. Huygens and one [Johannes] van Ceulen [watchmaker] about the invention and construction of very accurate timepieces, which would not deviate from each other by more than a second per day [24 hours], in which way the East and West could be found …

In view of the VOC’s encouragement, on 17 December 1683 Huygens sent his design drawings and an approximate model of his new invention to van Ceulen. Huygens seemed keen to test his novel (tricord) design in practice; his tricord clocks were pendulum clocks where the bob was attached to three strings suspended from a circular frame. Writing to the Dutch mathematician Bernhard Fullenius Jr., he stated that

> … at the request of the [East India] Company, I have undertaken the construction of clocks to determine the longitude, possessing as constant a regularity as those with the three-foot pendulum, but such as should not be disturbed by the motion of the sea. I found the task to be more difficult than I had initially thought, although it is not completed yet there is little doubt that it will succeed. (Huygens, 1683)

Van Ceulen soon produced two of the novel tricord clocks, which enabled Huygens to make a convincing case to the VOC, in July 1684, for financial and practical support. At their meeting on 27 July 1684, the governors of the VOC approved a financial contribution to van Ceulen so as to “complete the work to perfection,” recording that

> The Lord Mayor Hudde has stated that he is authorised by the resolutions of 31 December 1682 and 28 February 1684 to … spend one to two thousand guilders on [the endeavour] … despite not having achieved the aim … proposing whether the assembled [governors] could approve payment of two hundred silver ducats to the aforementioned van Ceulen, … on account of the work so far completed. (VOC governors, 1684)

On 13 August 1685, Huygens travelled to Amsterdam with his two new marine clocks, hoping to suspend and regulate them on board a VOC galleon. To his regret, however, “for the wind was perfect for a trial” (Huygens, 1685*a*), he did not manage to test them in practice, since Hudde was absent. Huygens’ next journey to Amsterdam, on 9 September 1685, was more fruitful. The VOC's governing board had assigned him a ship and a captain, Barent Fockes, at their meeting of 30 August 1685 (VOC governors, 1685); they also authorised Hudde to pay van Ceulen a second tranche of 200 ducats for his work and compensate the blacksmith he had retained with a fee of 70 guilders. Hudde was charged with retaining oversight, to make sure that the project would lead to 'enlightenment' of the state of maritime navigation.

Huygens met with a very courteous and amenable Barent Fockes, who had however been given instructions to sail to the VOC's main anchorage at the northern Dutch island of Texel rather than merely onto the Zuyderzee (Huygens, 1685*b*), the inland sea-turned-lake in present-day Netherlands. This would involve a 7–8 day voyage instead of the 2–3 days Huygens had anticipated, but the governors of the VOC insisted on a trial on the open sea. After all, Huygens wrote to his father, initial tests on open water would need to be successful, because "otherwise it would be useless to continue further afield."

On 11 September they embarked on their voyage from the port of Amsterdam, the only test ever undertaken by Huygens himself. However, because of a severe storm (Constantijn Huygens, 1685), the captain was forced to seek refuge at Enkhuizen Harbour, fearing damage to his sails. They completed their voyage to Texel a few days later. Despite the rough conditions, one of the clocks continued to run smoothly (Huygens, 1685*c*); the second clock ran intermittently. Nevertheless, Huygens was strengthened in his conviction that his clocks were seaworthy (Huygens, 1685*d*). He informed Hudde on 26 October 1685 that his envoy Thomas Helder was ready and prepared to take his marine clocks on an endurance voyage to the Cape of Good Hope (Huygens, 1685*e*). The VOC subsequently made provision for two of Huygens' clocks and two attendants to join its fleet on a voyage to the Cape.

Huygens prepared detailed instructions for Helder, his assistant Johannes de Graaff (de Graef) and their companion on the voyage, the clockmaker Willem van der Dussen, *Instructions and education as regards the use of the clock to find the longitudes of East and West*[22] (Huygens, 1686). His instructions included guidance regarding the mounting, regulating and maintenance of the timepieces.

The first VOC-sanctioned long-range voyage commenced on 24 May 1686, arriving at the Cape on 26 September. Much of the voyage was affected by rough seas, and so Helder did not manage to obtain any useful measurements. The East Indian return fleet set off from the Cape en route to Texel on 20 April 1687, but notes on the clocks' performance do not begin until 10 May 1687 (Huygens, 1687). When they returned to the home anchorage on 15 August 1687, Helder was no longer on board. He had died shortly after having left the Cape, in late April 1687, and many of his notes had disappeared. Huygens' second envoy, de Graaff, had taken over and managed to acquire enough measurements for Huygens to trace back the ship's course.

Huygens submitted his findings regarding his clocks' accuracy—*Report regarding longitude determination by clocks on the voyage from the Cape of Good Hope to Texel in the year 1687*[23]—to Hudde on 24 April 1688 (Huygens, 1688). He conceded that there was a small problem with the longitudes determined by his clocks: the measurements seemed to imply that the ship had sailed right through Ireland and Scotland: see Figure 14.

Huygens attributed the ship's apparent deviation from the VOC's navigators' route to the effects of the Earth's rotation: the 'spinning-off' (centrifugal) effect on bodies, and hence their loss of weight, was greater at the Equator than towards the Poles (for a discussion, see de Grijs, 2017). The updated route showed that they had clearly not sailed right through Ireland and Scotland on their way to the VOC's anchorage, and that the ship's terminus after 117 days at sea was just 19 km East of the actual longitude of Texel—for the time an unprecedentedly accurate determination of a ship's position at sea.[24] It amounted to a loss of just 68 seconds of clock time over the course of the voyage from the Cape to Texel.

It is unclear whether anyone not directly involved with the VOC or in the examination of Huygens' claims was aware of the significant discrepancies that had come to light when tracing back the ship's route based on the clocks' raw measurements. There is no evidence that Huygens commented in any correspondence on the apparent trajectory through Ireland and northern Scotland, although a memorandum by David Gregory, the inventor, from 11 November 1691 includes a passage which suggests that Huygens may have discussed the problem of the latitude dependence of the gravitational force during a journey to England in the summer of 1689:

> By observations of a ship from the Cape of Bonne Esperance [the Cape of Good Hope] to Texel on board which was a two of these Clocks, the course of the ship was on the coast of Ireland on the supposition the weight was the same in all parts of the earth or the pendulys vibration in equal times, but if the the [sic] other hypothesis of the less weight at the Equator be true the course will be (as it was) by the north of Scotland but both systems bring the ship to Texel. (*Gregory manuscripts*, 1627–1720)

In his letter to Hudde accompanying Huygens' report, he referred to another letter from the Dutch scholar and manuscript collector Isaäc Vossius to Coenraad van Beuningen, VOC governor. Vossius questioned the accuracy of the measurements, stating that …

> … the clock of Mr. Christiaen Huijgens [*sic*] performs excellently, but if one were to calibrate it based on the Eclipses, it will indicate during the 24 hours of a day and a night no more than 22 hours. (Vossius, 1688)

Huygens countered that in Vossius' letter,

> … where he objects to the observations of the Jesuits at the Cape of Good Hope and in general against observations of the longitude based on the Satellites of Jupiter, but both without any reason, since he has little knowledge of Astronomy and of the relevant type of observations … Because one cannot fathom what the meaning is of these words. (Huygens, 1688)

The close match of the end point of Huygens' corrected route with the independently known longitude of the Texel anchorage convinced him of the viability of his clocks as accurate marine timepieces. He summarised his conviction in the opening paragraph of his report to the VOC's governors:

> I can bring very good news concerning this invention, for I have found that by using the aforementioned clocks the longitudes between the Cape of Good Hope and Texel have on the whole been measured very well, and the total longitude between these two places [has been measured] so perfectly that it only deviates by 5 or 6 leagues, which I admit I have seen with exceptional satisfaction, it being certain proof of the possibility of this very-long-sought-after affair. (Huygens, 1688: Appendix II)

Huygens' reference to an accuracy of his positional determination of 5 or 6 leagues corresponds to approximately 30 km or 27 minutes of arc at the latitude of Texel. If Huygens' claim of such a high accuracy was indeed correct, this implies that he would have met the accuracy threshold (better than 30 minutes of arc, or 55 km at the Equator) for the award of the full prize money of £20,000 associated with the British Longitude Act of 1714. It is, therefore, curious that Huygens' report to Hudde was apparently never widely circulated among contemporary scientists nor translated into French, Latin or English. Perhaps the strength of Huygens' claim may not have been as great as implied by his bold assertion that his clocks could be used to determine the longitude without any reasonable doubt.

On 14 May 1689 Huygens' report and his route map were passed on to de Volder for formal review. The latter was generally supportive of Huygens' conclusions, but he cautioned the VOC governors that they should not place too much weight on the results of a single experiment (de Volder, 1689). His careful scrutiny of Huygens' analysis revealed that the Dutch scholar had made a mistake in his calculations pertaining to the ship's longitude on 8 June 1687, which also affected all subsequent calculations (de Volder, 1689: 341). The corrected values resulted in the improved accuracy of the ship's longitude with respect to that of the coast of Texel of 17 minutes of arc.

Despite the good agreement between the course recorded in the ship's log and Huygens' corrected trajectory, the latter was by necessity based on interpolation. In his report to the VOC, Huygens commented that …

> … the differences between the mariners and the corrected clocks are usually about 1 or 2 degrees, and always less than 3 degrees. And it should amaze no one that the mariners' reckoning would be 3 degrees off the true longitude on such a long voyage, because of the uncertainty in their guesses, from unknown currents and the ship's falling behind, as well as from its uncertain advancement. (Huygens, 1688: Appendix II)

Since this first test on board a VOC ship had left too many open questions, de Volder recommended that the VOC undertake a second sea trial. Once again, de Graaff was employed to take charge of the clocks. The outward leg commenced on 29 December 1690, arriving at the Cape on 4 June 1691. The return ship departed from the Cape on 26 June 1692, arriving at Veere Harbour, south of Rotterdam, on 10 October 1692.

Unfortunately, this second sea trial was anything but a success. First, only one of the two clocks taken on board was operational during the outbound voyage. Yet no useful data could be obtained during the first leg from Texel to the Cape Verdian port of St. Jago (São Tiago) because of inclement weather conditions at the anchorage that had prevented an initial calibration, despite repeated attempts to do so by de Graaff (1690*a*,*b*,*c*,*d*). The second clock was not operational at any time throughout the voyage (de Graaff, 1693; Huygens, 1693*a*).

On arrival at the Cape, de Graaff fell seriously ill for about three weeks. In addition, the need to obtain proper measurements and calibration of the clocks based on the Sun's motion, for which he needed at least three weeks, caused them to delay their return voyage by a full year (de Graaff, 1691; de Graaff et al., 1691). However, when he finally embarked on the return voyage, he did not install the clocks correctly, so that once again no useful data were obtained (Huygens, 1693*a*). Despite the array of errors affecting de Graaff's measurements (de Graaff, 1692), in his report Huygens (1693*a*) significantly understated the rather disastrous results, reporting that "the clocks have not proved such a success as we had hoped for."

In addition to issues related to the accuracy of the measurements, the trial also suffered from a number of basic flaws in the experimental set-up. For instance, no attempt was made to check Huygens' corrections independently, either by obtaining measurements with a 'seconds pendulum' or by calibrating the measurements using the longitudes of well-known landmarks. The latter could have been ascertained by observations of the ephemerides of Jupiter's moons at the Cape. Indeed, this was one of Huygens' main recommendations in his final report:

> It would still be very helpful if one investigated the true longitude at some important places with regard to the Meridian of Texel or Amsterdam, by observing the satellites of Jupiter. (Huygens, 1688: Appendix II)

In fact, Huygens specifically suggested that measurements taken on both legs of the voyage *at the same place* should be compared with the accurately known timing of Jupiter's satellites to provide the final proof. Unfortunately and to Huygens' significant dismay, his recommendation was ignored (Huygens, 1693*c*). Nevertheless, Huygens' suggestion to use the satellites of Jupiter for calibration purposes brings his efforts to achieve an accurate determination of longitude at sea back to the scholar's original suggestion: use of Jupiter's moons had been suggested from the outset (Huygens, 1658). Practical considerations called for more straightforward timing measurements, however.

The accuracy required for the trial to be deemed successful had been clearly communicated. Initially, a VOC resolution of 31 December 1682 called for an accuracy of better than one second deviation per 24 hours. This was subsequently relaxed to a performance requirement of better than two seconds per 24 hours in a new resolution dated 28 April 1684. In a letter of 24 March 1693, Huygens (1693*c*) commented to de Volder on the accuracy of the observations taken during the second voyage. He concluded that a positional deviation of 10 minutes 51 seconds, corresponding to a time error of 43.4 seconds, had been incurred over a period of six days, or a time loss of 7.2 seconds per day.

Huygens eventually conceded that the jury was still out on the performance of his clocks:

> I do not want to pretend that one could conclude from this or based on the previous trial of A°. 1687 that the perfection [of my method of] Longitude measurement has been demonstrated conclusively. (Huygens, 1693*d*)

However, he was keen that de Volder put in a good word for him with the VOC governors (Huygens, 1693*d*). It has been suggested (Schliesser and Smith, 2000) that his concession regarding his clocks' accuracy was driven by his ongoing development of a new marine clock design, the *balancier marin parfait* (perfect marine balance), which he wanted to start promoting shortly. This is supported by his final comments in a letter to the VOC governors of 6 March 1693 (Huygens, 1693*b*), which he repeated in a note to de Volder of 19 April 1693 (Huygens, 1693*d*), hinting at forthcoming developments:

> I have on this occasion invented something quite different and incomparably better, which I have in hand at the present moment, whereby any little difficulty in the use of this invention will once and for all be removed, of which in due course I hope to give Your Excellencies further particulars, ...

Huygens had started work on a radically new type of clock around the time of the first sea trial of his pendulums on the Zuyderzee in 1684 (Huygens, 1683–1684), which eventually led to his design of the perfect marine balance. The need for a reliable clock to determine longitude at sea remained unabated. Huygens was keen to test his new device in practice (Huygens, 1683–1684), but death intervened. He passed away on 8 July 1695, before being able to achieve his lifelong goal of manufacturing a sufficiently accurate timepiece for use at sea. Details of Huygens' perfect marine balance are scant, given that he was still working on its design at the time of his death.

He had derived a new curve that would ensure isochronous operation of a clock without being adversely affected by the rocking and pitching motions of ships at sea. His basic premise was that that the pendulum's driving force had to be proportional, at any time, to the extent the pendulum is out of equilibrium. His new isochronous regulator consisted of a vertical balance wheel equipped with a chain that linked two systems of small, equal, and equidistant weights. While swinging, each of these systems would rise and fall, and thus they would alternately increase and decrease the force exerted on the chain by the weights. The resulting torque acting on the wheel was directly proportional to its displacement from equilibrium, thus leading to isochronous oscillations (for details, see de Grijs, 2017: 5-49–5-51).

Huygens expected the chain to oscillate slowly, even on choppy seas, so that the effects of the latter would be greatly reduced. Initial tests showed that the device did not operate satisfactorily, however: large oscillations were completed more slowly than expected. Huygens attributed this behaviour to a combination of increased air resistance associated with larger oscillations and the large number of weights that had to be kept going while their angles with respect to the vertical direction were changing constantly (see de Grijs, 2017: 5-49–5-51).

## 6 AFTERMATH AND FINAL THOUGHTS

Whereas the Dutch longitude competition remained open for submissions, by the middle of the seventeenth century the initial flurry of applications had slowed to a mere trickle. Much of the continuing effort was coordinated centrally, first by Galileo's supporters and subsequently by Huygens and his associates. Nevertheless, non-allied proposals did find their way to the States General or the States of Holland fairly regularly. One particularly noteworthy episode in the Dutch longitude quest at this time descended into a rather unpleasant polemic, which eventually did not result in much progress towards the ultimate goal.

From the 1680s, Bernhard Fullenius 'de Jongere' (junior), professor at the University of Franeker (now defunct), corresponded increasingly frequently with Huygens in the context of their efforts to find a solution to the longitude problem. However, their constructive engagement was soon rudely interrupted by the appearance of one Lieuwe Willemsz (Graaf), a Mennonite priest and a loud-mouthed, belligerent Frisian mathematician and almanac composer. Over the course of the next decade, Fullenius and Willemsz became embroiled in an increasingly fierce and acrimonious polemic (Dijkstra, 2012).

Willemsz soon associated himself with and looked for guidance to Matthias Wasmuth, a controversial German professor in Hebrew at the academy in Rostock, who would subsequently be appointed as a professor of divinity at the University of Kiel. Towards the end of his life, he turned his focus and interests to astronomy, because he believed that his insights into the main religious texts he was familiar with allowed him to make a major breakthrough (Bayerische Akademie Der Wissenschaften, 2018). He claimed that divine revelation had facilitated him to perfect his breakthrough, which in turn made him act as if he was the new leader of the European astronomical community.

Willemsz had devised his own ideas on longitude determination at sea (Graaf, 1689: 4–5), and he sought the assistance of the publicist Balthasar Bekker to help him disseminate his ideas more widely. Bekker admitted that Willemsz' work was too complicated for his understanding, so he recommended that the latter seek input from his brother-in-law, Fullenius Jr. (Bekker, 1692: 19–20) A meeting was arranged in 1688, but when Fullenius asked Willemsz to explain the foundations of his calculations, Willemsz refused because he did not trust Fullenius' motives, fearing that his secret would be stolen (Davids, 1995). As a direct consequence of his perceived mistreatment by Fullenius, Willemsz started a slanderous campaign against the former (Fullenius, 1689), while simultaneously drumming up support for his method from powerful patrons across the province of Friesland (Dijkstra, 2007). Fullenius maintained that he did not think that Willemsz idea would work in practice, which thus gave rise to a fierce polemic that became increasingly political in nature (for full details, see Dijkstra, 2012).

Willemsz was eventually allowed to appear before the Deputy States of Friesland, who offered him a letter of recommendation to present his ideas on solving the longitude problem to a commission of the States General (Davids, 1986: 131, 426). Nevertheless, Willemsz never fully disclosed in detail how he intended to solve the longitude problem. He only discussed his method in vague terms; if and when he revealed any details, he did so in the form of examples or 'proofs' (his designation). He claimed that he had been inspired by both a divine revelation and the ideas of Wasmuth, which had allowed him to calculate ephemeris tables. To determine one's longitude, it would be sufficient to measure the distance between the moon and one of a number of designated reference stars. His tables could then be used to determine one's position on Earth. Whereas this type of lunar distance method had become well-established by the end of the seventeenth century (e.g., de Grijs 2020*a*), Willemsz claim was less than trustworthy given that he claimed to be able to calculate the notoriously uncertain lunar orbit to unprecedented precision—without disclosing how.

In addition, Willemsz' tables purportedly also provided a direct correlation between the lunar distance measured on the sky and the distance on Earth between the observer and the biblical Garden of Eden. In essence, his method was said to be useful for calculating the local time in the Garden of Eden (Graaf, 1691), but without the need to know the local time at the observer's position, a physical impossibility as the method required just a single free parameter to find the longitude. Willemsz claimed to have discovered a 'big secret'—which he however never disclosed—allowing him to calculate the exact position of the Garden of Eden (Dijkstra, 2012). Eventually, his contemporaries turned away from this loud-mouthed miscreant, who was too precious about his ideas to disseminate them more widely.

Individual requests to be considered for any of the Dutch longitude prizes, if only for the expenses component, continued to trickle in until well into the eighteenth century. In the Delft city archives (2020), we find records of hopeful projectors filing patent applications for the determination of longitude at sea from as recently as the 1740s and 1750s. These include a submission by one Pieter de Fay, accountant in Amsterdam, from 1746–1750 and another application filed in 1756 by one G. G. Stokman from Isleben, near Leipzig.

De Fay had filed a patent application with both the provincial States of Holland, and with the States of Zeeland, for his method to determine “the length of East and West at sea” as early as 19 August 1746, and again on 31 December of that year. On 14 March 1749, the States of Holland retained Pieter van Musschenbroek and Johan Lulofs, professors of philosophy at Leiden University, to examine de Fay’s method. As a result, on 26 March 1749, the States of Holland resolved that de Fay’s method was neither novel nor sufficient to warrant a patent (States of Holland, 1749). In a second resolution issued on 27 February 1750, the States of Holland agreed to ask the deputies of the city of Delft to re-examine the proposed method so as to determine whether the applicant could be awarded compensation for the expenses incurred in the development of his method and the construction of the relevant devices (States of Holland, 1750).

Stokman’s request for a subsidy and a patent arrived at the States of Holland on 22 February 1756. His request was discussed at the next council meeting on 28 February, where it was decided to request an expert assessment from the deputies of the city of Delft as part of a committee convened by the States of Holland (States of Holland, 1756).

The second half of the eighteenth century saw further institutionalised attempts to secure means to determine longitude at sea. As we saw already, in 1787 the Admiralty of Amsterdam established its Longitude Committee. This was a clear sign that state agencies were enthusiastic about substantially increasing their involvement in finding suitable practical means in support of the main challenges in navigation. The first Longitude Committee was composed of Jan Hendrik van Swinden, professor of mathematics, physics, astronomy, logic and metaphysics at the Amsterdam *Athenaeum Illustre*; Pieter Nieuwland, van Swinden’s protégé; Gerard Hulst van Keulen, the leading publicist of anything nautical and an expert teacher of longitude methods; and—from 1789—Jacob Florijn, representative of the Admiralty of Rotterdam.

In the National Archives of the States of Holland, a detailed treatise from 1793 by Hendrik de Hartog(h), mathematician, navigator and astronomer at the Amsterdam *Athenaeum Illustre*, is listed (States of Holland Archives, 2020). His work covers longitude determination and is identified as a follow-up publication to one published around 1771 by Pybo Steenstra, one of his predecessors at the *Athenaeum Illustre*. De Hartog is said to also have been involved in the early developments of the Longitude Commission (Zandvliet, 1999). In 1787, he published his first manuscript dealing with longitude determination based on lunar distances (de Hartog, 1793), which became the mandated method for ships of both the Admiralty of Amsterdam and the VOC in 1788. The record in the National Archives refers to a complete treatise on this subject, which earned him a promotion to examiner of the VOC on 22 February 1790. For a more in-depth discussion of the Longitude Committee’s remit and achievements until it dissolution in 1850, I refer the reader to Davids’ (2015) lucid essay.

It should have become clear that Dutch efforts to find a practically viable solution to the longitude problem were as rich as those put forward to the Spanish Crown around the same time, as well as those submitted in response to the establishment of the British Longitude Prize in 1714. Whereas most overviews of Dutch efforts focus on Huygens’ pursuit of a practical marine timepiece, and sometimes also on Galileo’s proposals to use the ephemerides of Jupiter’s satellites, comprehensive reviews of the full series of developments during this period are lacking, even in Dutch. The present article aims to remedy that situation, while at the same time showcasing the breadth of ideas originating from the tolerant republic on the North Sea, from methods based on magnetic declinations and lunar distance measurements to the development of stable marine timepieces.

The start of the Scientific Revolution, exemplified by Galileo’s astronomical ‘heresy’, combined with increased trade across the world’s oceans created the ideal conditions across the European continent to make significant progress on the pre-eminent scientific and practical problem of the era, the need for a viable method of longitude determination at sea.

## 7 NOTES

[1] On 15 February 1600, the States General had received a letter from Prince Maurits dated 10 February 1600, informing the governing body that one Jacob van Straten claimed to have "found the height from the East and the West"; little else is known about him, other than that he seems to have jealously guarded his secretive solution to the longitude problem (Davidse, 2000–2020).
[2] The *stuiver* was a pre-decimal Dutch coin; twenty *stuivers* represented a monetary value equal to one guilder or—as indicated in this quotation—one pound.
[3] Galileo Galilei is usually simply referred to by his first name, a convention I have adopted here too.
[4] The Dutch surname 'van Velsen' is rather common. A detailed search of online archives yielded only a certain 'Willen van Velsen' as possible identification for this person; see the *Resolutiën Staten-Generaal 1576–1625* (see Note 8) and the *Repertorium van ambtsdragers en ambtenaren 1428–1861* (http://resources.huygens.knaw.nl/repertoriumambtsdragersambtenaren1428-1861).
[5] Figure 3 is a portrait gallery of the main characters driving the developments described in this article, if and when their images were available in the public domain. I encourage the reader to refer to this compilation whenever a new character voicing an original solution is introduced in a more than cursory manner.
[6] Like Figure 3, Figure 4 is a portrait gallery of officials, examiners and other important support personnel who were instrumental in driving the longitude discussion forward.
[7] *Commissie tot de Zaaken, het Bepalen der Lengte op Zee en het Verbeteren der Zeekaarten Betreffende.*
[8] http://resources.huygens.knaw.nl/retroboeken/statengeneraal/.
[9] *Een Tractaet seer dienstelijck voor alle Zee-varende Luyden/door het t'samen spreken van twee Piloten.*
[10] *Het Ghebruyck der naeld wiisinge tot dienste der zee-vaert beschreven.*
[11] *Klare vertooninge, hoe men door het uurwerk van Elohim ofte den grooten Eyghenaer, namelijck door son, maan en sterren in alle plaetsen der werrelt zijn meridiaenslenghte sal kunnen vinden.*
[12] *De grouwelijke ongehoorde blasphemien ende raserijen van Thomas Leamer.*
[13] *Grondelijcke Onderwijsinghe van de Sterrekonst.*
[14] *Nyeuwe taeffelen der Polus-hoochte enz.*
[15] He is often confused with his namesake who went to Sweden to become a key economic adviser to Kings Karl IX and Gustav II Adolphus and their governments; it appears, however, that this identification is incorrect (Molhuysen et al., 1927).
[16] *Reken-Konst Vande Groote See-vaert Waer doormen op alle streken van't Compass so wel de Lenghde Oost ende West kan vinden als de Breede Zuydt ende Noort, Ghepractiseert, beschreven ende int licht ghebracht to dienst van alle Zeevarende Lieden.*
[17] *Het gulden zeeghel des grooten zeevaerts, daerinne beschreven wordt de waerachtige grondt vande zeylstreken en platte pas-caerten (voor desen noyt bekent) waermetals onder een secreten Zeegaende ghenerale Regule vant gesicht des groote Zeevaerts bevesticht en tot zyn vollencomen perfectie gebracht wordt, dienende tot een Voorlooper van de voorschreven Regule, daerinne mede ghesien worden, de onbehoorlycke Procedueren die de wedersprekers seghen deselve ende het ghemeene best drie jaren lanck hebben ghepleecht.*
[18] *Voyage vant experiment vanden generalen regul des Gesichts vande groote zeevaert.*
[19] As a case in point, the online *Rijks Geschiedkundige Publicatiën* (National Historic Publications) contain a witness statement from 17 September 1621 in favour of Jarichs van der Ley's method (Dutch National Historic Publications, 1510–1672).
[20] *Discorsi e dimostrazioni matematiche intorno a due nuove scienze.*
[21] This probably refers to an earlier version of the second edition of Placentinus' *Geotomia.*
[22] *Instructie en onderwijs aangaende het gebruijk der Horologiën tot het vinden der Lengde van Oost en West.*
[23] *Rapport aengaende de Lengdevindingh door mijne Horologiën op de Reys van de Caep de B. Esperance tot Texel A° 1687.*
[24] The discrepancy originally reported by Huygens was 25 minutes of arc; de Volder's corrections led to the adjusted value adopted here.

## 8 REFERENCES

The following abbreviation is used: *HOC* = *Oeuvres Complètes de Christiaan Huygens.*